\documentclass[a4paper,11pt]{article}
\usepackage{jinstpub} 
\usepackage{lineno}
\usepackage{siunitx}
\usepackage{circuitikz}
\usepackage{mathtools}
\usepackage{amsmath}
\usepackage[toc,page]{appendix}
\usepackage{comment}

\DeclareSIUnit {\atmosphere}{atm}
\DeclareSIUnit {\curie}{Ci}
\DeclareSIUnit {\torr}{torr}
\DeclareSIUnit {\year}{yr}

\title{\boldmath Project 8 Apparatus for Cyclotron Radiation Emission Spectroscopy with $^\mathrm{83m}$Kr and Tritium}

\author{
A.~Ashtari Esfahani$^1$,
D.M.~Asner$^2$,
S.~B\"oser$^3$,
N.~Buzinsky$^4$,
R.~Cervantes$^1$,
C.~Claessens$^{1,3}$,
L.~de~Viveiros$^5$,
P.~J.~Doe$^1$,
J.~L.~Fernandes$^2$,
M.~Fertl$^3$,
J.~A.~Formaggio$^4$,
D.~Furse$^4$,
L.~Gladstone$^6$,
M.~Guigue$^7$,
J.~Hartse$^1$,
K.~M.~Heeger$^8$,
X.~Huyan$^2$,
A.~M.~Jones$^2$,
J.~A.~Kofron$^1$
B.~H.~LaRoque$^2$,
A.~Lindman$^3$,
E.~Machado$^1$,
E.~L.~McBride$^1$,
P.~Mohanmurthy$^4$,
R.~Mohiuddin$^6$,
B.~Monreal$^6$,
E.~C.~Morrison$^2$,
J.~A.~Nikkel$^8$,
E.~Novitski$^1$,
N.~S.~Oblath$^2$,
M.~Ottiger$^1$,
W.~Pettus$^9$,
R.~Reimann$^3$,
R.~G.~H.~Robertson$^1$,
G.~Rybka$^1$,
D.~Rysewyk$^4$,
L.~Salda\~na$^8$,
P.~L.~Slocum$^8$,
M.~G.~Sternberg$^1$,
Y.-H.~Sun$^6$,
P.~T.~Surukuchi$^{8,10}$,
J.~R.~Tedeschi$^2$,
A.~B.~Telles$^8$,
F.~Thomas$^3$,
M.~Thomas$^2$,
T.~Th\"ummler$^{11}$,
L.~Tvrznikova$^{12}$,
B.~A.~VanDevender$^2$,
M.~Wachtendonk$^1$,
M.~Walter$^{11}$,
J.~Weintroub$^{13}$,
T.~E.~Weiss$^{4,8}$,
T.~Wendler$^5$,
N.~L.~Woods$^1$,
E.~Zayas$^4$,
A.~Ziegler$^5$
}

\affiliation{$^1$ Center for Experimental Nuclear Physics and Astrophysics and Department of Physics, University of Washington, Seattle, WA 98195, USA\\
$^2$ Pacific Northwest National Laboratory, Richland, WA 99354, USA\\
$^3$ Institute for Physics, Johannes Gutenberg University Mainz, 55128 Mainz, Germany\\
$^4$ Laboratory for Nuclear Science, Massachusetts Institute of Technology, Cambridge, MA 02139, USA\\
$^5$ Department of Physics, Pennsylvania State University, University Park, PA 16802, USA\\
$^6$ Department of Physics, Case Western Reserve University, Cleveland, OH 44106, USA\\
$^7$ Laboratoire de Physique Nucléaire et de Hautes Energies (LPNHE), Sorbonne Université, Université Paris Diderot, CNRS/IN2P3, 75005 Paris, France\\
$^8$ Wright Laboratory, Department of Physics, Yale University, New Haven, CT 06520, USA\\
$^9$ Center for Exploration of Energy and Matter and Department of Physics, Indiana University, Bloomington, IN, 47405, USA\\
$^{10}$ Department of Physics \& Astronomy, University of Pittsburgh, Pittsburgh, PA 15260, USA\\
$^{11}$ Institute for Astroparticle Physics, Karlsruhe Institute of Technology, 76021 Karlsruhe, Germany\\
$^{12}$ Lawrence Livermore National Laboratory, Livermore, CA 94550, USA\\
$^{13}$ Center for Astrophysics | Harvard \& Smithsonian, Cambridge, MA 02138, USA
}

\emailAdd{ashtari@uw.edu}

\abstract{
Cyclotron Radiation Emission Spectroscopy (CRES) is a novel technique for the precise measurement of relativistic electron energy.
This technique is being employed by the Project~8 collaboration for measuring a high-precision tritium beta decay spectrum to perform a frequency-based measurement of the neutrino mass.
In this work, we describe the Project 8 Phase II apparatus, used for the detection of the CRES signal from the conversion electrons of $\mathrm{^{83m}Kr}$ and the first CRES measurement of the beta-decay spectrum of molecular tritium.
}


\begin{document}
\maketitle
\flushbottom

\section{Introduction}
The observation of neutrino flavor transformation proved that neutrinos have non-zero mass~\cite{super-k, SNO}, a result that contradicts the expectation from the Standard Model of Particle Physics. 
However, the absolute mass scale of these fundamental fermions remains unknown~\cite{ParticleDataGroup:2024cfk}.

Various techniques are being used in the ongoing quest to measure neutrino mass.
If neutrinos are Majorana particles, their mass can be measured in the laboratory by searching for neutrinoless double-beta decay \cite{KamLAND-Zen:2024eml, GERDA:2020xhi, Majorana:2022udl, EXO-200:2019rkq, CUORE:2021mvw}. 
Cosmological measurements of large-scale structure, interpreted in the $\Lambda$CDM model framework, offer another indication of the neutrino mass scale, albeit with significant model-dependence \cite{plank,eboss,desi}. 

The kinematics of beta decay \cite{KATRIN2024, PRL} and electron capture \cite{Holmes, ECHo} are affected by finite neutrino masses that cause distortions of the decay spectrum which provide direct and model-independent access to the neutrino masses, independently of their Dirac or Majorana nature~\cite{Drexlin:2013lha, Formaggio:2021nfz}.
These experiments are sensitive to an electron-weighted neutrino mass
\begin{equation}
    m_\mathrm{\beta} = \sqrt{\sum_{i=1}^3 |U_{ei}|^2 m_i^2},
\end{equation}
where $U_{ei}$ are elements of the Pontecorvo–Maki–Nakagawa–Sakata matrix and $m_i$ are neutrino mass eigenvalues \cite{Pontecorvo, mns}.

The current state-of-the-art direct neutrino mass experiment is KATRIN, which has an ultimate goal of sensitivity to a (anti)neutrino mass of \SI[per-mode = symbol]{300}{\milli\electronvolt}$/c^2$ (90\% C.L.) \cite{katrin-jphysg, KATRIN2024}.
KATRIN measures the beta-decay spectrum of molecular tritium, whose shape near the endpoint is modified by the neutrino mass.
However, investigating the mass range below \SI[per-mode = symbol]{100}{\milli\electronvolt}$/c^2$ with tritium is only practically possible in atomic form to avoid the energy broadening caused by the rovibrational initial and final states of the molecules \cite{molecular-t-3, molecular-t-2, Formaggio:2021nfz}.
Extending the neutrino mass reach of beta-decay experiments also requires excellent energy resolution, low background, high statistics, and careful control of systematic effects.

The Project 8 collaboration has pioneered the Cyclotron Radiation Emission Spectroscopy (CRES) technique, targeting a next-generation direct (anti)neutrino mass sensitivity of \SI[per-mode = symbol]{40}{\milli\electronvolt}$/c^2$ using tritium beta decay.
This technique measures the kinetic energy of an electron via the frequency of the cyclotron radiation it emits while undergoing cyclotron motion in a magnetic field.
The cyclotron frequency is
\begin{equation}\label{cres-eq}
    f_\mathrm{c} = \frac{f_\mathrm{0}}{\gamma} = \frac{1}{2 \pi} \frac{|e|B}{m_{\mathrm{e}} + K_{\mathrm{e}}/c^2} \ ,
\end{equation}
where $f_\mathrm{0}$ is the non-relativistic cyclotron frequency, $\gamma = 1 + K_{\mathrm{e}}/m_{\mathrm{e}} c^2$ is the Lorentz factor, $B$ is the magnetic field, and $e$, $m_{\mathrm{e}}$, and $K_{\mathrm{e}}$ are the electron's charge, mass, and kinetic energy \cite{CRES-2009}. While neutrino mass measurement is a primary application of Cyclotron Radiation Emission Spectroscopy, it is also used to measure highly relativistic electron and positron energies from $\mathrm{^{6}}$He and $\mathrm{^{19}}$Ne beta decays, opening a new avenue for searches for signatures of beyond Standard Model physics through precision beta-decay measurements~\cite{He-6, He-6-2}.
There is also a proposal to utilize CRES for ultra-high resolution X-ray spectroscopy~\cite{x-ray}.

The Project~8 collaboration plans to reach its design sensitivity to neutrino mass as the conclusion of a program in four phases \cite{whitepaper, CRES-2017}.
Phase I demonstrated CRES in a small active volume using a rectangular waveguide to collect the cyclotron radiation from mono-energetic conversion electrons emitted by metastable krypton atoms ($\mathrm{^{83m}}$Kr) \cite{CRES-2015}.
In Phase II, described here, the first tritium spectrum was measured in an upgraded but still-small ($\mathcal{O}$(cm$^3$)) circular-waveguide-based apparatus, using the same magnet.
This measurement resulted in the extraction of the first frequency-based neutrino mass limit of $m_\beta < $ \SI{155}{\electronvolt}$/c^2$ (\SI{152}{\electronvolt}$/c^2$) in a Bayesian (frequentist) analysis~\cite{PRL, PRC}.
Phase III will demonstrate the key technologies for a full-scale measurement: the production and trapping of cold atomic tritium, and CRES in large ($\mathcal{O}$(m$^3$)) volumes using a resonant cavity detector with higher efficiency and improved resolution. 
The final Phase IV will scale to reach the ultimate neutrino mass sensitivity goal of \SI{40}{\milli\electronvolt}$/c^2$ \cite{bayesian}.

\begin{figure}
  \centering
  \includegraphics[width=140mm]{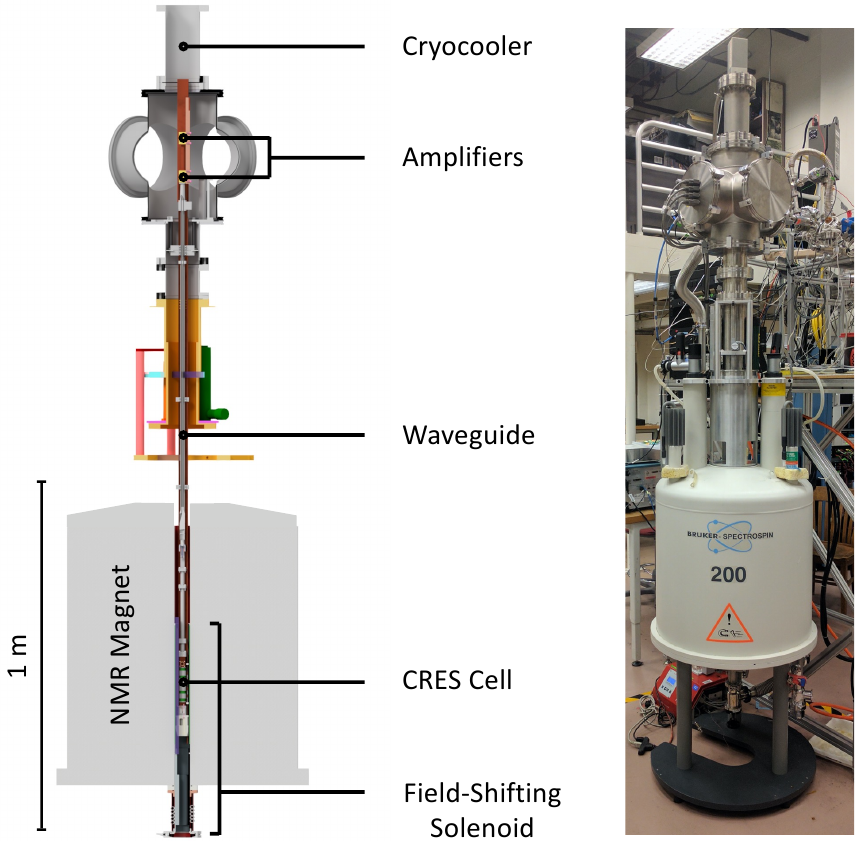}
  \caption{A model cross-section (left) and photo (right) of the apparatus used in Phase II of the Project~8 experiment.
    The superconducting NMR magnet is visible at the center.
    The CRES cell was supported by a cryogenic insert, suspended inside the field-shifting solenoid within the magnet's bore.
    Two cryogenic amplifiers attached to the cryocooler fixed the signal to noise ratio of the entire signal chain.
  }
  \label{ch3:magnet}
\end{figure}

Phase II of Project~8, in addition to setting the first CRES-based neutrino mass limit, was a crucial demonstration of CRES’s potential as a precision measurement technique~\cite{PRL,PRC}.
Achieving these goals relied on the control and precise characterization of the Phase II apparatus. There were zero counts above the tritium endpoint, setting a stringent background rate limit and demonstrating that RF noise, the primary source of potential background, could be understood and excluded.
Built-in calibration tools made it possible to map the relationship of detection efficiency with signal frequency, allowing for the control of related uncertainties.
The leading contributors to detector response\textemdash the magnetic trap's field profile, and effects due to scatters of electrons off gas molecules and atoms\textemdash were also characterized precisely, leading to a 1.66-eV FWHM resolution in $\mathrm{^{83m}}$Kr data.
In this work, we describe the design and characterization of the apparatus (Fig.~\ref{ch3:magnet}) that enabled these advances~\cite{ashtari}.

In this Phase II experiment, the cyclotron frequency is $\sim$\SI{26}{\giga\hertz} for the highest-energy electrons in the tritium spectrum in a \SI{0.959}{\tesla} magnetic field. A Nuclear Magnetic Resonance (NMR) magnet was used to produce the field (Sec. \ref{ch2:magnetl}).
A cryogenic insert positions the CRES cell, a waveguide section where the decay electrons are produced and measured, at the maximum of the magnetic field.  The insert also supports the cyclotron radiation transfer to the amplifiers, the radioactive gas connection, and the coils to generate the magnetic field for electron trapping (Sec. \ref{ch3:insert}).
After amplification, a room-temperature receiver system filtered, down-converted, and digitized the signal (Sec. \ref{ch4:receiver}).
The last essential part of the Phase II apparatus was the gas delivery system, which was designed to provide the proper composition and pressure of the radioactive gases (Sec. \ref{ch5:Gas-system}).

\section{Magnet}\label{ch2:magnetl}
A warm-bore Bruker-Spectrospin \SI{200}{\mega\hertz}
superconducting NbTi NMR magnet (Fig.~\ref{ch3:magnet}) produced the \SI{0.959}{\tesla} magnetic field of the experiment.
The magnetic field was pointed upwards, towards the amplifiers, resulting in left-handed circularly polarized radiation.

\begin{figure}
  \centering
  \includegraphics[width=120mm]{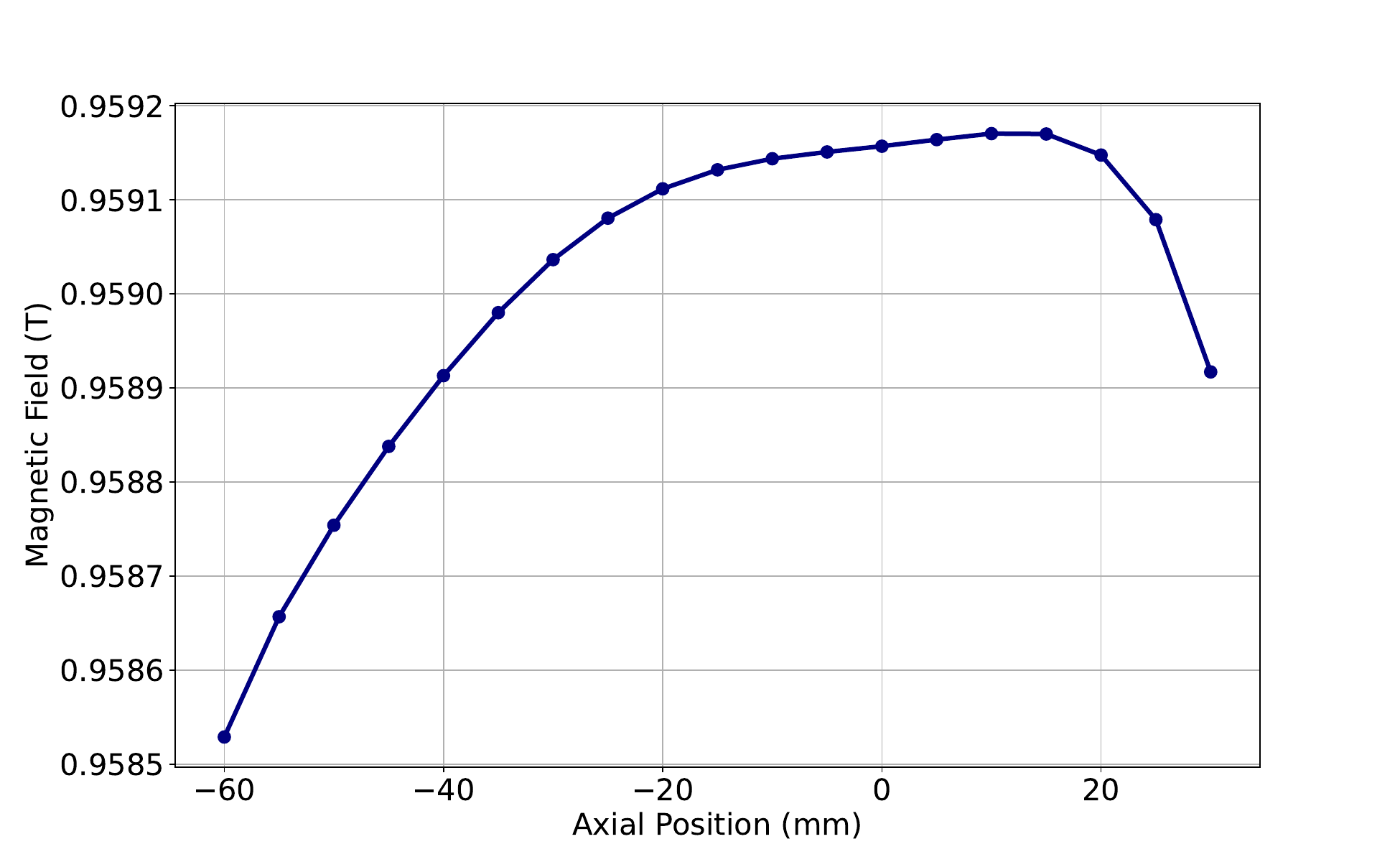}
  \caption{The magnetic field measurement at the center of the magnet bore on the axis of the magnet, performed in April 2018.
    The insert's axial position corresponds with the choice where the physics data was recorded. 
    The zero of the axial position is at the center of the central trap coil.
  }
  \label{ch3:nmr-magnet}
\end{figure}

A non-functional trim coil limited the magnetic field homogeneity, but the 10 ppm homogeneity achieved across the \SI{52}{\milli\meter} diameter of the bore with the seven working shim coils was sufficient for Phase II of Project 8 \cite{jared}.
Fig.~\ref{ch3:nmr-magnet} shows an on-axis magnetic field strength measurement with a Metrolab PT2025 NMR teslameter.
A Pfeiffer HiCube 80 turbomolecular pump was used to keep the pressure in the bore below \SI{10}{\micro\torr}.
The CRES cell outer vacuum system was connected to the magnet bore by a sliding piston seal and could be raised and lowered by means of a hydraulic cylinder.
The CRES cell vertical position in the background field was adjusted to maximize field homogeneity across the electron trapping volume.
The desired vertical position was then fixed with machined aluminum blocks.

\begin{figure}
  \centering
  \includegraphics[width=120mm]{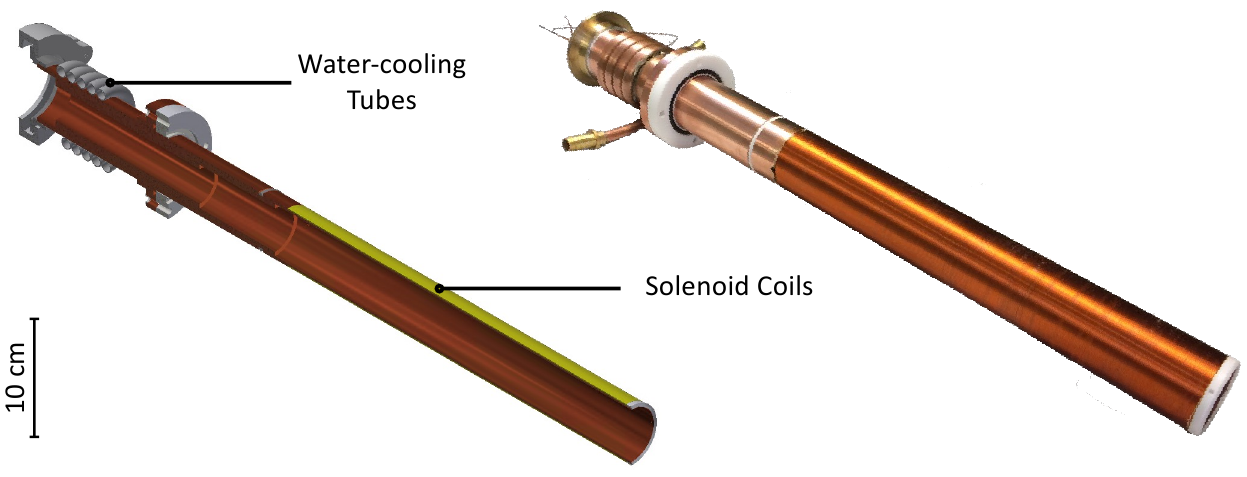}
  \caption{A model cross-section (left) and photo (right) of the field-shifting solenoid used in Phase II of the experiment.
    The solenoid was \SI{350}{\milli\meter} long with an outer diameter of \SI{52}{\milli\meter} and an inner diameter of \SI{46}{\milli\meter}. 
    It was used to homogeneously shift the background magnetic field to vary the cyclotron frequency of mono-energetic electrons from $\mathrm{^{83m}}$Kr for studying frequency-dependent systematic effects.}
  \label{ch3:fss-coil}
\end{figure}

A non-superconducting solenoid, called the field-shifting solenoid, was installed inside the NMR magnet's bore to alter the magnetic field at the location of the CRES cell for systematic studies (Fig.~\ref{ch3:fss-coil}) \cite{PRC, Christine}.
The background field was shifted to vary the cyclotron frequency of \SI{17.824}{\kilo\electronvolt} mono-energetic conversion electrons from $\mathrm{^{83m}}$Kr to investigate the frequency-dependent features of the signal.
Two layers of AWG26 copper magnet wire potted in varnish were wound onto a \SI{1.4}{\milli\meter} thick copper cylinder that fit around the cryogenic insert, within the NMR magnet.
The coil form was brazed to a bolted O-ring flange that mated to the bottom of the NMR magnet, below which were brazed five rounds of copper tubing for cooling water.
A \SI{1}{\ampere} current generated a \SI{4.9}{\milli\tesla} field, which was sufficient to shift the cyclotron frequency of the \SI{17.824}{\kilo\electronvolt} electrons by $\sim$\SI{130}{\mega\hertz}.
Using COMSOL simulation software, the spatial variation along the CRES cell's length in the field created by the field-shifting solenoid was estimated to be less than \SI{40}{\micro\tesla}.

The temperature of the CRES cell was initially stable with the field-shifting solenoid installed, but later became unstable due to the development of a poorly-understood thermal link.
Since the frequency-dependent systematic studies had already been concluded by then, the solenoid was removed before proceeding with the tritium data campaign in Phase II.

\section{Cryogenic Insert}\label{ch3:insert}
\begin{figure}
  \centering
  \vspace*{20pt}
  \includegraphics[width=150mm]{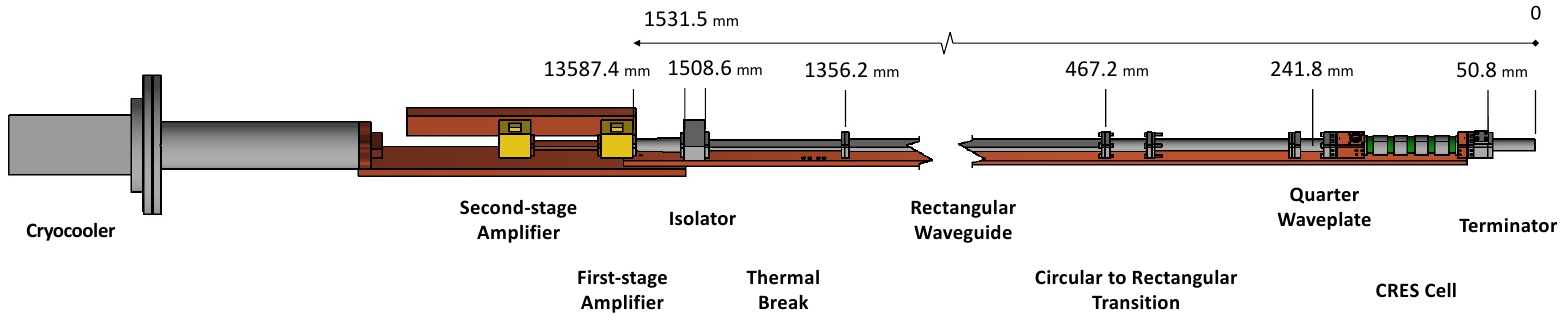}
  \caption{Cryogenic insert.
    The CRES cell is shown on the left side of the picture.
    Multiple components of the RF power transfer section are apparent (see Sec. \ref{power_transfer}).
    The cryocooler is connected to the cryogenic insert on the right side of the picture.
  }
  \label{insert}
\end{figure}

The combination of the CRES cell, the RF power transfer section, and the amplifier stage comprised the cryogenic insert. (Fig. \ref{insert}).
A \SI{2.5}{\centi\meter} $\times$ \SI{1}{\centi\meter} copper bar was the mechanical backbone for the cryogenic insert.
The copper bar and the amplifiers were connected in parallel to a Cryomech AL60 single-stage Gifford-McMahon cryocooler using copper braids.
This group of components was held at low temperatures to minimize the addition of thermal noise before the amplification of the CRES signal.
The amplifiers were at $\sim$\SI{30}{\kelvin}, while the CRES cell temperature was raised to \SI{85}{\kelvin} using a software PID (proportional-integral-differential) loop-stabilized heater to prevent the $\mathrm{^{83m}}$Kr from freezing to the CRES cell walls.
This temperature was experimentally optimized to maximize the recorded event rate \cite{PRC}.

\subsection{CRES Cell}\label{Crescell}
\begin{figure}
  \centering
  \includegraphics[width=120mm]{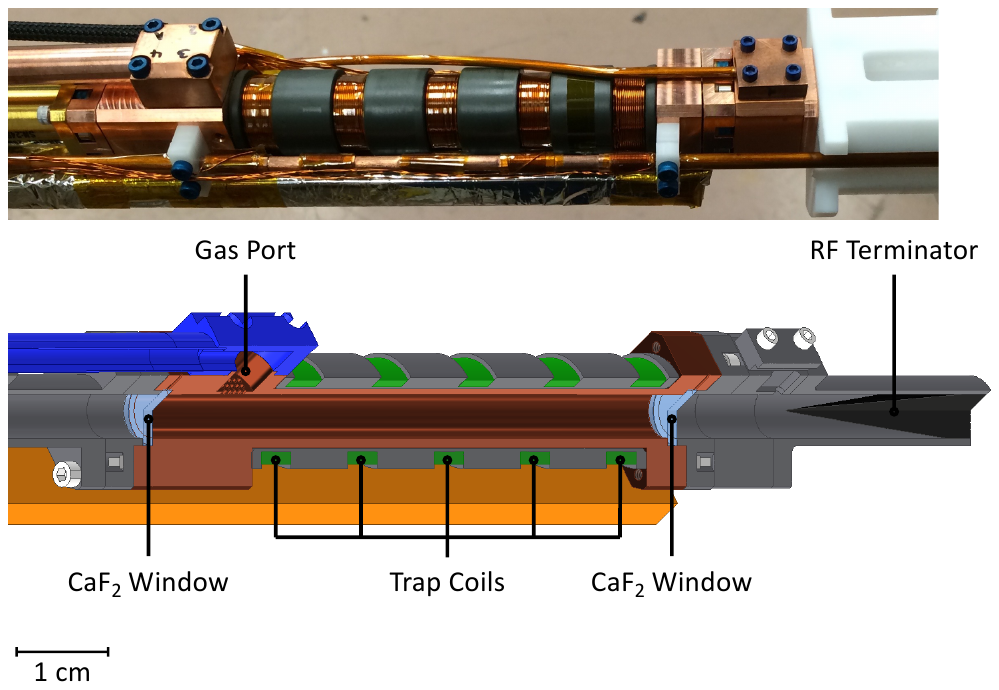}
  \caption{The CRES cell used in Phase II of the experiment (top) and its schematic diagram (bottom).
    Five trapping coils were available to generate the desired magnetic trapping geometry.
    A Torlon support structure connected the traps, thermally and mechanically, to the CRES cell.
    Two $\mathrm{CaF_2}$ windows confined the radioactive gas inside the cell.
    Radioactive gases were admitted through a dedicated gas port.
    The cone-shaped terminator, manufactured with epoxy and graphite, was used to reduce the effects of interference.
  }
  \label{ch3:gascell-2}
\end{figure}

The CRES cell (Fig. \ref{ch3:gascell-2}) was designed to be an efficient microwave guide for the frequency range of interest.
Furthermore, the cell had to be able to confine the radioactive gas while disturbing the waveguide's RF properties as little as possible.
Therefore, two microwave-transparent windows were installed into the waveguide structure.

Two \SI{2.4}{\milli\meter} thick pieces of $\mathrm{CaF_2}$ were selected for the construction of the windows.
This length corresponds to half a wavelength at \SI{24.5}{\giga\hertz}\footnote{$\mathrm{CaF_2}$ has a relative permittivity of 6.5 at temperatures lower than \SI{100}{\kelvin} \cite{caf2-e}.}.
RF simulations confirmed the windows' transparency in the frequency region of interest.
Furthermore, the similarity of the thermal expansion coefficients of $\mathrm{CaF_2}$ and copper ensured a safe cooldown where the vacuum integrity of the CRES cell was preserved to avoid a tritium leak \cite{caf2}.
Laser-cut indium gaskets were used to make the vacuum seal between the windows and the cylindrical cell wall.
Titanium and aluminum bolts sustained mechanical pressure over the entire range from room temperature to \SI{50}{\kelvin}, preserving vacuum integrity.

\begin{figure}
  \centering
  \includegraphics[width=120mm]{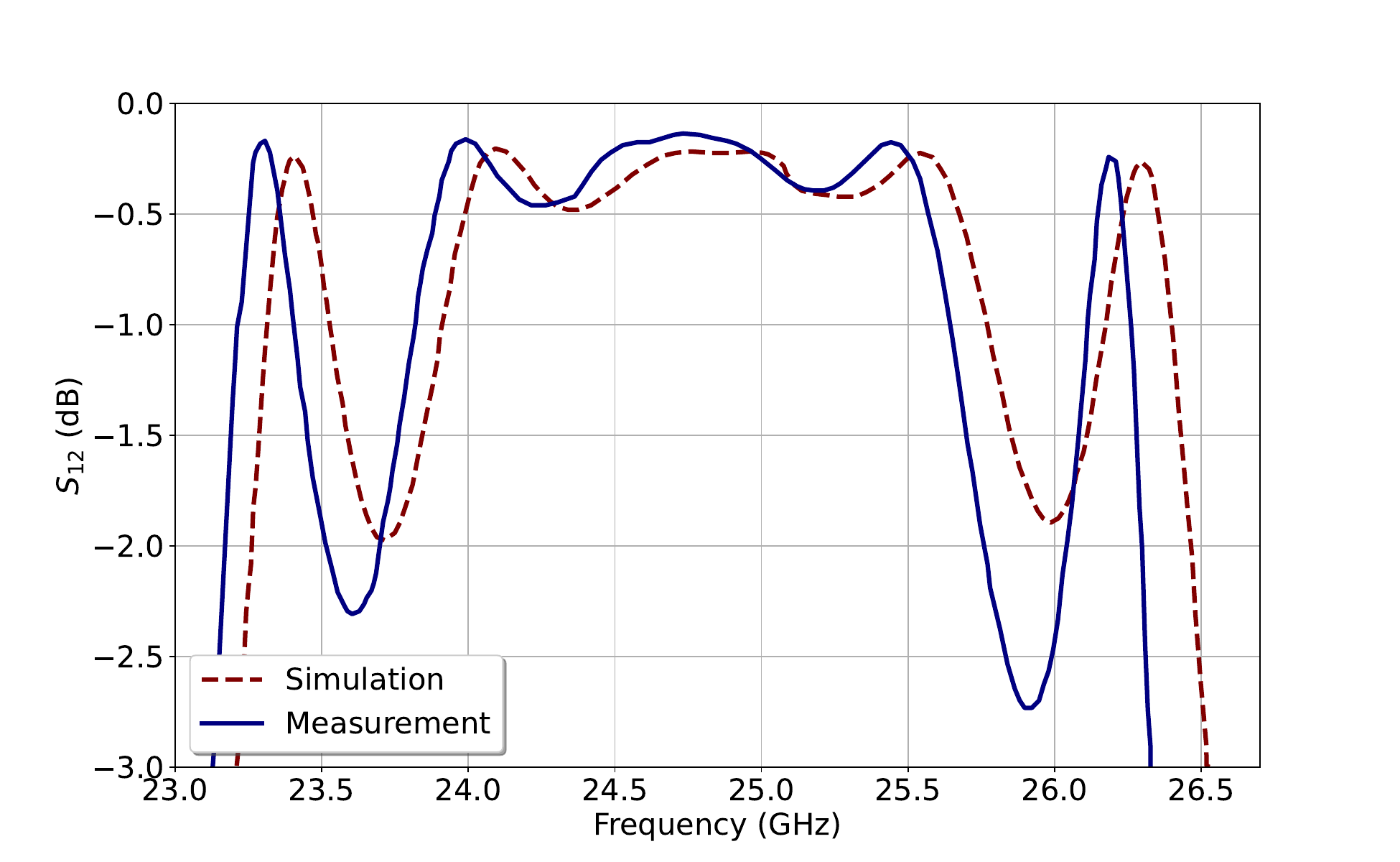}
  \caption{Simulated and measured transmission performance for the entire CRES cell assembly with $\mathrm{CaF_2}$ windows installed at room temperature. The 0.4\% frequency shift is accredited to differences in the $\mathrm{CaF_2}$ window thickness and dielectric constant in simulation vs. assembled cell.
  }
  \label{ch3:windows-rf}
\end{figure}

After installing the $\mathrm{CaF_2}$ windows, the RF transmission properties of the CRES cell were measured with a network analyzer.
The results shown in Fig.~\ref{ch3:windows-rf} confirmed a \SI{1.7}{\giga\hertz} wide frequency window with $\leq$ \SI{0.5}{\decibel} attenuation.
However, not all of the bandwidth was available, as the signal was down-converted by \SI{24.5}{\giga\hertz} (Sec. \ref{ch4:receiver}).

\subsection{Magnetic Trap}\label{ch3:magnetic-bottle}

The strong homogeneous magnetic field that causes cyclotron motion confines electrons radially. However, without axial confinement, relativistic electrons would escape the observation region in a fraction of a microsecond, making it impossible to detect the CRES signal over noise. A magnetic bottle was used for axial trapping.

Five coils made of AWG 26 copper wire were wound around the microwave guide on a Torlon support structure to produce the magnetic field for electron confinement (Fig.~\ref{ch3:gascell-2}).
The high thermal conductivity of Torlon enabled sufficient cooling for coils operating at \SI{1}{\ampere}.
The coils were also potted in GE varnish 7031 to facilitate heat transfer out of the coils.

\begin{figure}
  \centering
  \includegraphics[width=120mm]{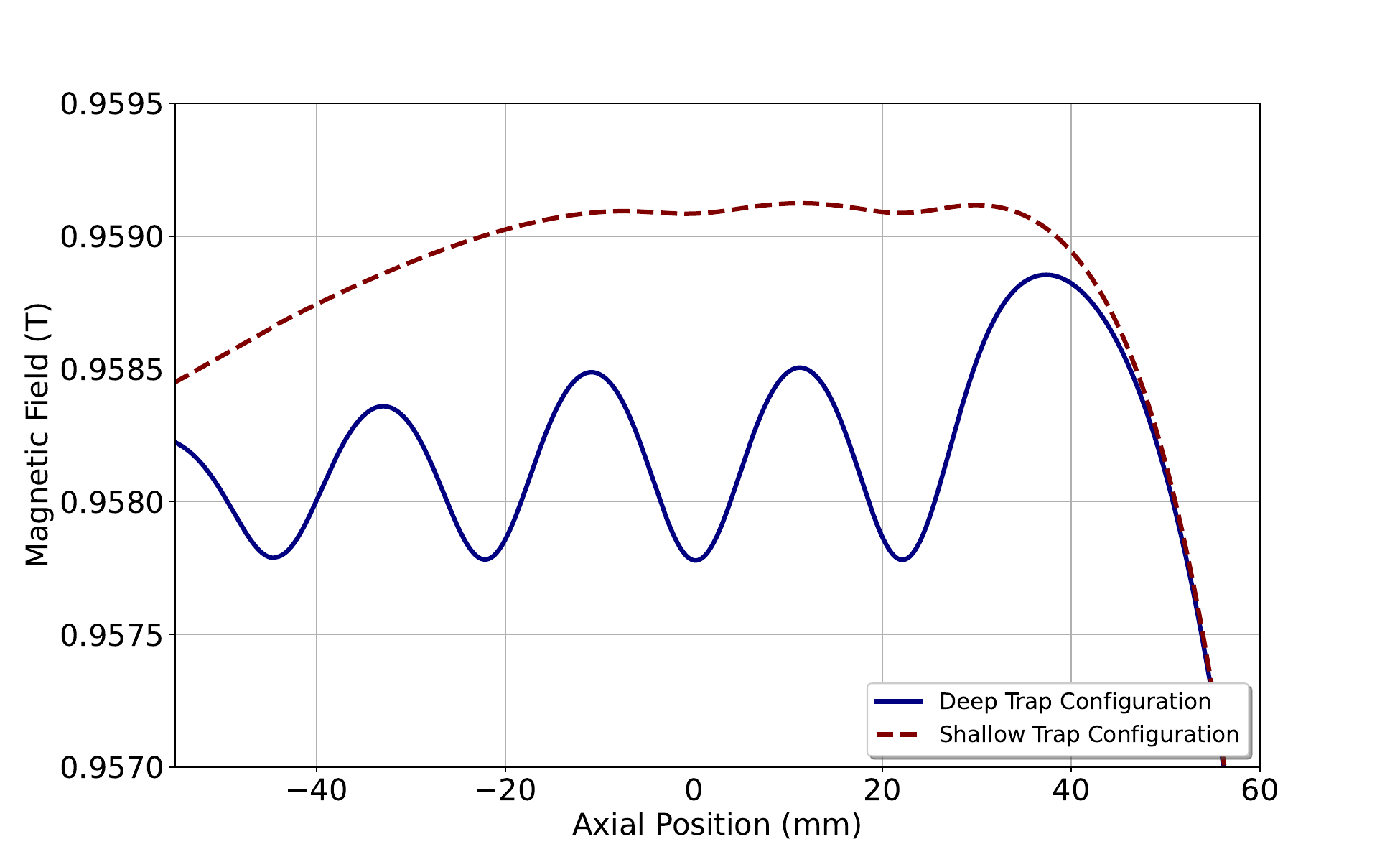}
  \caption{The magnetic field for the two common trap configurations used in Phase II.
    The overall shape of the field is formed by the background field (Fig.~\ref{ch3:nmr-magnet}) while the local minima are generated by the magnetic field of the traps.
  }
  \label{trap_config}
\end{figure}

Two different trap configurations were used in Phase II.
First, to establish the high-resolution capabilities of the CRES technique, a "shallow trap" configuration with two trap coils was used with small currents (\SI{12.00}{\milli\ampere} and \SI{17.85}{\milli\ampere}, supplied by two Stanford Research Systems LDC 501 precision power supplies) to minimize the systematic effects caused by the magnetic field variation experienced by the trapped electrons (Fig.~\ref{trap_config}).
The two currents differed to compensate for the inhomogeneity in the background field.
Second, for acquiring data at higher event rates, a "deep trap" configuration with four field dips (\SI{174}{\milli\ampere}, \SI{249}{\milli\ampere}, \SI{272}{\milli\ampere}, and \SI{286}{\milli\ampere}, with the additional currents supplied by an AIM-TTI MX100TP power supply) was used.
The deep-trap configuration increased the event rate by a factor of 40 at the expense of a factor-of-30 worsening in the energy resolution.
The fifth coil was not used in this configuration as it was sited in the steep region of the background field, making it so that even the highest current that could be used in this trapping coil without overheating did not create a local minimum in the magnetic field.
Only the deep-trap configuration was used for the tritium spectrum measurement because because the neutrino mass uncertainty was statistics-limited, whereas both shallow- and deep-trap configurations were used for krypton measurements~\cite{PRL, PRC}.

\subsection{Waveguide Design}
\begin{figure}
  \centering
  \includegraphics[width=120mm]{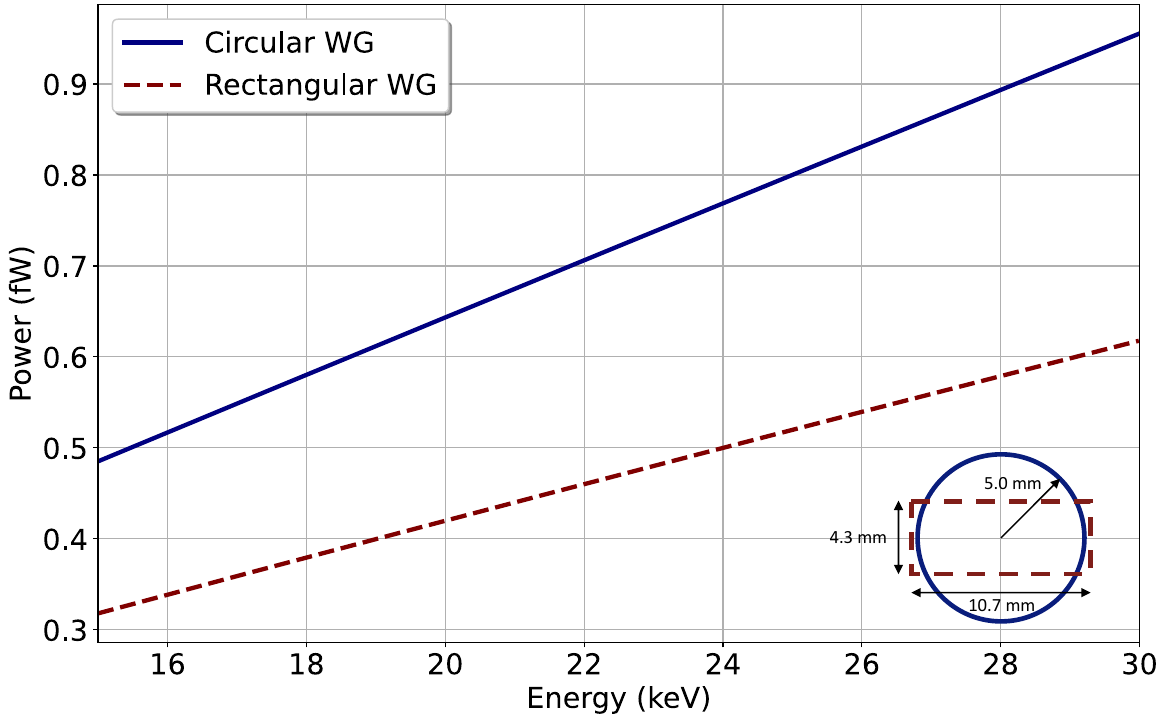}
  \caption{Electron radiated power into the fundamental modes of the circular waveguide used in Phase II and the WR-42 rectangular waveguide used in Phase I of the experiment, as a function of electron energy.
    The calculation is based on the phenomenological model developed in \cite{pheno-19}.
    The cross-sections of the waveguides are shown in the lower right corner.}
  \label{ch3:rect-circ}
\end{figure}

The waveguide design in Project 8 was optimized for electron energies in the range of \SIrange[range-units = single]{10}{32}{\kilo\electronvolt}.
These electrons radiate power in the microwave K-band frequency range in the presence of the \SI{0.959}{\tesla} magnetic field.
A segment of a K-band circular waveguide, with an inner diameter of \SI{10}{\milli\meter} (0.396"), was used as the CRES cell in Phase II of the experiment.\footnote{Increasing the volume of the electron trapping region would increased the number of radioactive gas molecules and the number of decays observed. 
However, increasing the cross-sectional dimensions could have induced coupling of the electrons to non-fundamental modes in the waveguide, increasing the complexity of data analysis.}
Electrons' cyclotron motion couples to the first two modes of the circular waveguide, $\mathrm{TE_{11}}$ and $\mathrm{TM_{01}}$.
The symmetry of a cylindrical waveguide with perpendicular electric field components in its fundamental mode ($\mathrm{TE_{11}}$) provides a more efficient coupling to the cyclotron electron compared with the fundamental mode of a rectangular waveguide ($\mathrm{TE_{10}}$) used in Phase I of the experiment.
Fig.~\ref{ch3:rect-circ} compares the power radiated into the fundamental modes for electrons in the circular CRES cell used in Phase II with that for the rectangular waveguide used in Phase I.
With its larger cross-section, the circular waveguide also had a larger physical volume and therefore higher radioactivity in the cell at a given pressure and temperature.
The choice of copper for the CRES cell limited the attenuation due to the resistive losses in the waveguide walls to be less than \SI{0.25}{\decibel~\meter^{-1}} in the frequency range of interest.
Fig.~\ref{ch3:gascell-2} shows the CRES cell for Phase II.
The gas port connection to the CRES cell was through a grid of sub-wavelength holes, which did not degrade RF transmission along the signal path.

\subsection{RF Power Transfer and Amplification} \label{power_transfer}
The RF power transfer section of the apparatus was designed to allow the microwave radiation to propagate from the electron in the CRES cell to the cryogenic amplifiers (Fig. \ref{insert}).
This section's length was defined by the need to locate the isolator and the two low-noise amplifiers outside the strong magnetic field region.

After leaving the CRES cell, circularly polarized radiation was converted to linear polarization in a QuinStar MN: QWL–26MKFOZ396 left-circular polarization quarter-wave plate.
Next, a circular-to-rectangular waveguide adapter transitioned the electromagnetic radiation into the \SI{0.9}{\meter} long copper WR-42 waveguide, which carried the radiation out of the strong field region.

\begin{figure}
  \centering
  \includegraphics[width=120mm]{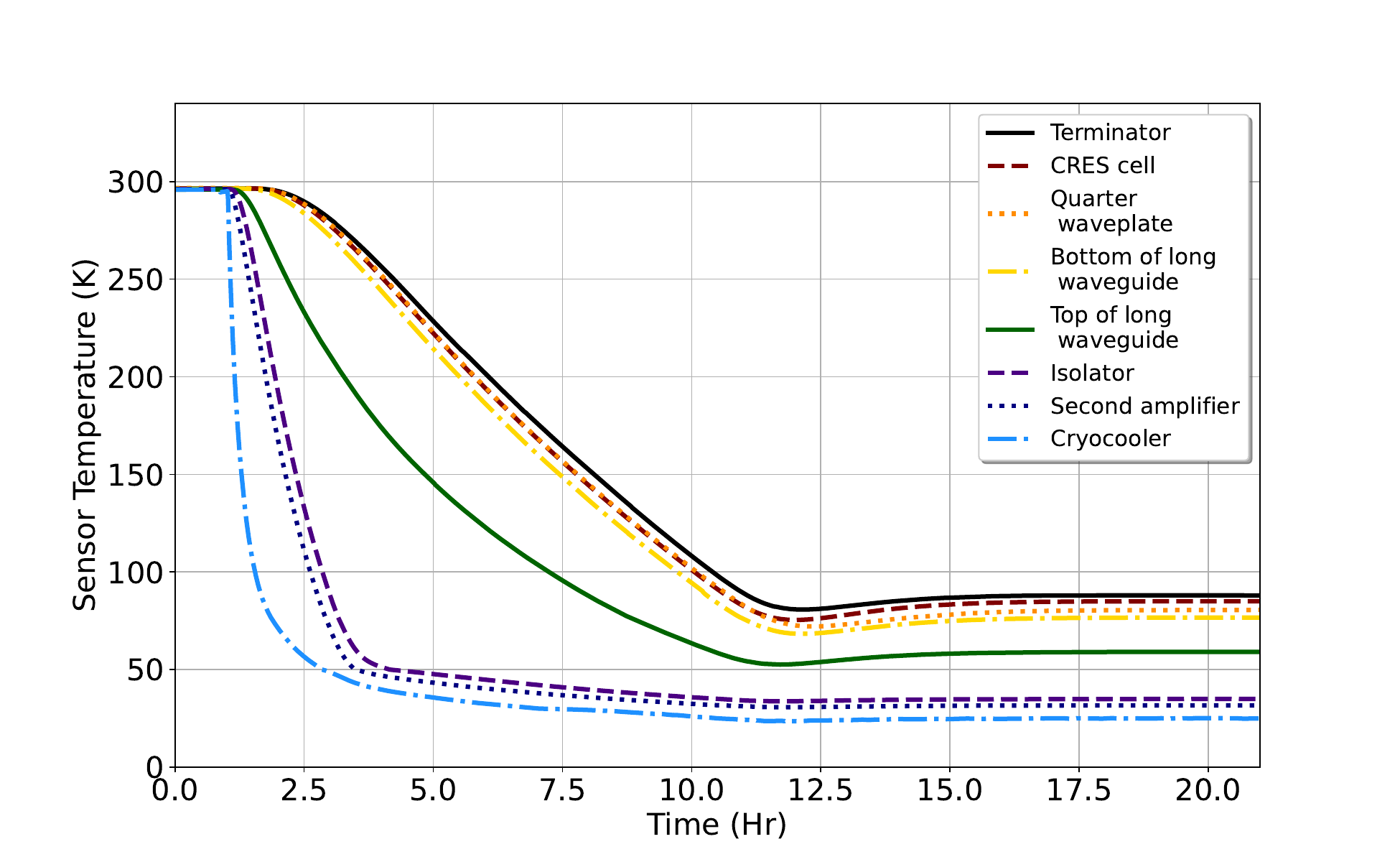}
  \caption{Temperature profile along the insert during a system cool-down.
    First, the temperature drops following cryocooler activation.
    Then, the CRES cell heater is activated to raise and stabilize CRES cell at its \SI{85}{\kelvin} operating temperature.}
  \label{cooldown_temp}
\end{figure}

To prevent the heat from the \SI{85}{\kelvin} region from heating the amplifiers (which would have raised the noise temperature), a low-thermal-conductivity but also low-RF-loss gold-coated stainless steel section of WR-42 waveguide, \SI{152}{\milli\meter} long, was added following the long copper WR-42 waveguide.
Fig. \ref{cooldown_temp} shows the temperature profile along the insert during a system cool-down, demonstrating the effectiveness of the thermal break in isolating the temperature of the amplifiers and isolator from the \SI{85}{\kelvin} CRES cell.
A Raditek RADI-18-26.5-Cryo-(4-77K)-WR42 RF isolator (a circulator with one port terminated) then terminated any power reflected from the amplifiers and flattened the frequency dependence of the gain and noise.
Finally, a waveguide taper transitioned to the WR-28 input port of the first of a cascaded pair of \SI{28}{\decibel} gain Low Noise Factory LNF-LNC22-40WA amplifiers, which were connected by a section of gold-coated WR-28 waveguide.

\subsection{Waveguide Terminator}
The CRES radiation propagating toward the lower end of the CRES cell was absorbed by a custom-made cone-shaped terminator.
Although an RF reflector would have increased the signal-to-noise ratio, interference of the reflected and the propagated waves would have generated frequency and position-dependent signal features\footnote{As was observed in Phase I, and in Phase II before the installation of the terminator.} \cite{pheno-19}.

Commercial cryogenic RF terminators were not available in the required geometry, so a terminator was fabricated locally.
It was cast from a non-magnetic and RF-dissipative mixture of Stycast 1266 epoxy and superfine natural graphite spherical powder. 
A conical design was chosen to provide a gradual change in impedance that maximally suppressed reflections.

\begin{figure}
  \centering
  \includegraphics[width=120mm]{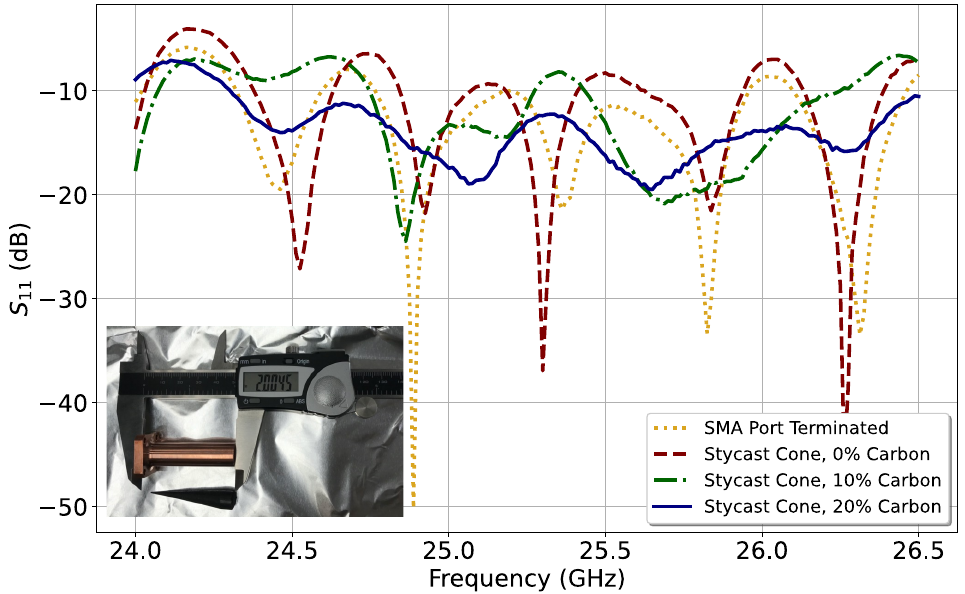}
  \caption{Measured reflection coefficients for the three locally fabricated epoxy/graphite terminators and for a commercial ANNE-50L+12 GHz SMA terminator.
    The commercial terminator was imperfect in this frequency range because it was designed for \SI{12}{\giga\hertz}.
    Ultimately, the terminator with $20\%$ carbon was used.
    The epoxy/graphite cone and waveguide housing that made up the terminator are shown at the bottom left.
  }
  \label{ch3:terminator-test}
\end{figure}

Three terminators with $0\%$, $10\%$, and $20\%$ graphite concentration by weight were cast and characterized.
Fig.~\ref{ch3:terminator-test} shows the reflection coefficients of these terminators compared to a commercial terminator.
The terminator with $20\%$ graphite concentration was selected because its attenuation was the least frequency-dependent.

\section{RF Receiver System}\label{ch4:receiver}

\begin{figure}
  \centering
  \includegraphics[width=140mm]{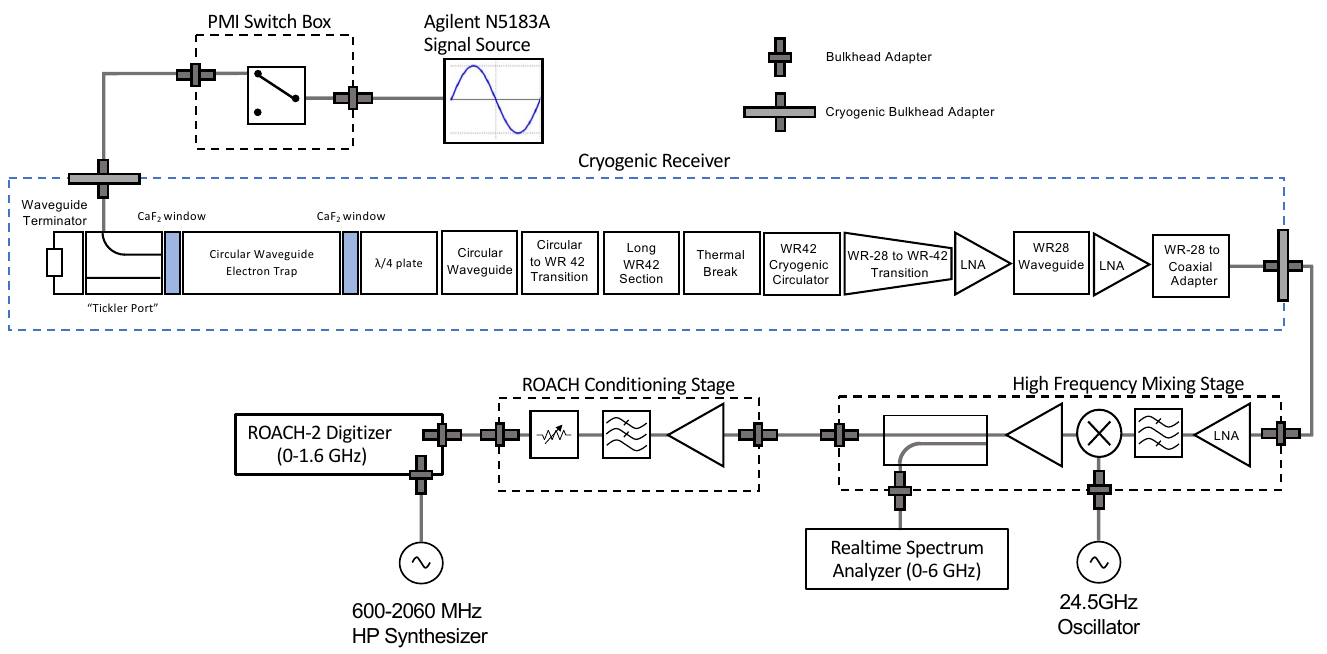}
  \caption{The Phase II RF receiver chain.
    The signal-to-noise ratio was fixed in the cryogenic receiver following the second low-noise amplifier.
    A high-frequency mixing stage bandpass filtered the signal and mixed down to intermediate frequency (IF).
    A second conditioning stage provided additional filtering and amplification.
    Two digitizer systems were available: primary physics data was collected with a ROACH-2 digitizer system, while a Realtime Spectrum Analyzer (RSA) was used for system commissioning and for validation.}
  \label{fig:phase2-receiver}
\end{figure}

After the cold signal amplification (see Sec.~\ref{power_transfer}), a series of room-temperature RF components were used to filter, downmix, amplify, digitize, and monitor the CRES signals (see Fig.~\ref{fig:phase2-receiver}). 
Here, these warm RF components are detailed, followed by the result of an analysis of the gain and the noise of the RF receiver chain.
A common \SI{10}{\mega\hertz} reference signal from a Stanford Research Systems FS725 rubidium clock was provided throughout the RF system.

\begin{figure}
  \centering
  \includegraphics[width=\textwidth]{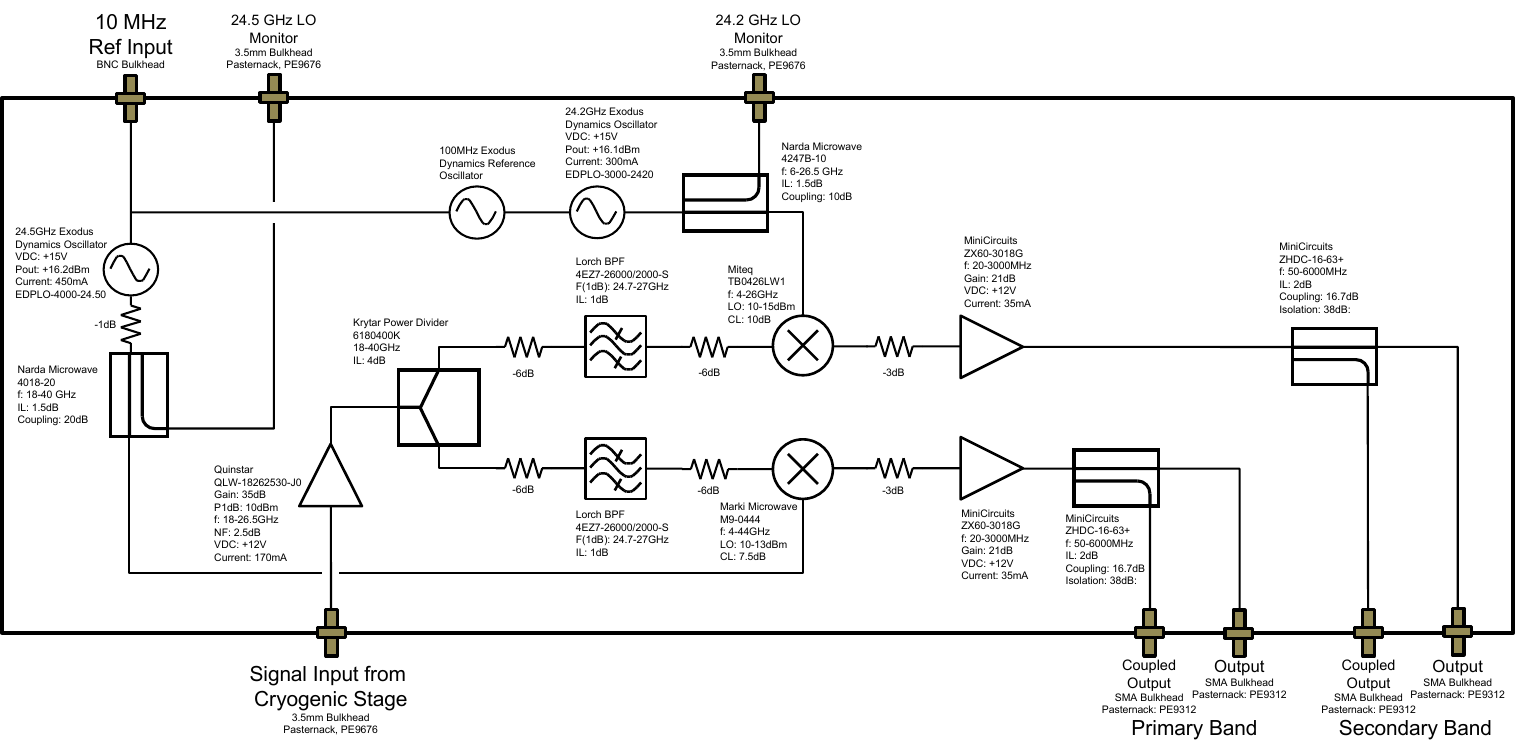}
  \caption{The Phase II high-frequency mixing stage.
    The $\sim$\SI{26}{\giga\hertz} RF CRES signals were bandpass-filtered and downmixed to IF by \SI{24.5}{\giga\hertz} for the primary band.
    The output and coupled outputs were connected to the ROACH conditioning stage and RSA, respectively.}
  \label{fig:hf-stage}
\end{figure}

The high-frequency mixing stage was designed to convert the raw RF signals down to intermediate frequency (IF) $\lesssim$\SI{2}{\giga\hertz}, suitable for digitization (see Fig.~\ref{fig:hf-stage}).
The input signal was carried on a short coaxial cable from the vacuum space feed-through to this stage.
The signal was first amplified by a low-noise Quinstar QLW-18262530-J0 amplifier before a Krytar 6180400K matched-line directional two-way power divider split the signal into two paths.
Both signal paths used a Lorch 4EZ7-26000/2000-S bandpass filter with a range 24.7\,--\,\SI{27.0}{\giga\hertz} to avoid image bandwidth noise when mixed.
The primary signal path was downmixed with a \SI{24.5}{\giga\hertz} reference from an Exodus Dynamics EDPLO-4000 phased-locked oscillator using a Marki Microwave M9-0444 double-balanced mixer.
A secondary signal path was downmixed with a \SI{24.2}{\giga\hertz} reference and comparable components.
Attenuators and amplifiers along the RF path ensured the signal was maintained in the optimal range and noise was minimized.
Directional couplers allowed for the monitoring of the oscillator signals or mixed signals from the high-frequency mixing stage in parallel with data taking.

System commissioning was performed with signals downmixed by the secondary \SI{24.2}{\giga\hertz} oscillator, available from the Phase I receiver.
These measurements determined \SI{24.5}{\giga\hertz} to be the ideal reference, and the high-frequency stage was rebuilt as depicted to accommodate the parallel primary and secondary bands, allowing a direct crosscheck of performance with the commissioning setup.
This reference choice provides full coverage of the tritium endpoint region within the Reconfigurable Open Architecture Computing Hardware-2 (ROACH-2) digitizer~\cite{roach} sampling bandwidth, placing the tritium endpoint (\SI{18.574}{\kilo\electronvolt}) at \SI{1.371}{\giga\hertz} IF.
The increased mix frequency limited the Lorch filter noise rejection at the low frequency end; the noise power increase from image noise was measured to be \SI{0.5}{\decibel} at \SI{0.2}{\giga\hertz} IF.
The transfer function of the high-frequency mixing stage was directly measured and also modeled with a cascade analysis, and good agreement was found.
The gain was 16\,--\,\SI{22}{\decibel} peaked at \SI{0.4}{\giga\hertz} IF; the modest gain falloff is due at low frequency to the Lorch filter response and at high frequency to the amplifier gain response.

The coupled output of the primary IF band was connected to a Tektronix RSA5106B Real-Time Signal Analyzer.
The RSA served as the primary data acquisition device during system commissioning owing to its wider tunable signal frequency bandwidth and dynamic range, as well as its established performance from Phase I.

The ROACH digitizer~\cite{Hickish2016}, which was integrated into a more flexible data acquisition system~\cite{DAQ}, served as the production digitizer for all Phase II physics data~\cite{Christine}.
The ROACH FPGA was clocked at \SI{1.6}{\giga\hertz} by a signal from a Hewlett-Packard 8648D signal generator.
The ROACH uses an 8-bit EV8AQ160 quad ADC chip, which digitizes interleaved at \SI{3.2}{\giga S/s} for a Nyquist frequency of \SI{1.6}{\giga\hertz}.
The ROACH input was directly digitized, requiring a conditioning stage to appropriately match the input.

\begin{figure}
  \centering
  \includegraphics[width=100mm]{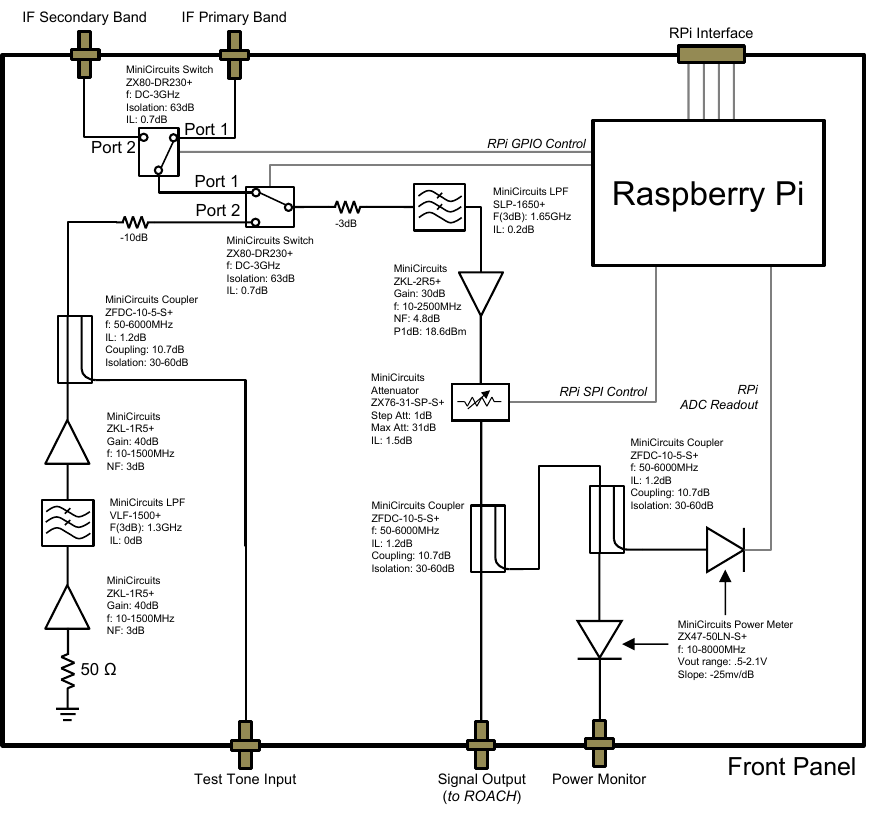}
  \caption{The Phase II ROACH conditioning stage.
    IF CRES signals from the high-frequency mixing stage were low-pass filtered and amplified to optimally match the ROACH-2 digitizer system.
    An integrated Raspberry Pi controlled the ROACH input source and attenuation level, as well as regularly monitoring the power level output to the ROACH.}
  \label{fig:roach}
\end{figure}

The high-frequency mixing stage outputs were connected by long cables to the ROACH conditioning stage (see Fig.~\ref{fig:roach}).
The signal was first low-pass filtered with a MiniCircuits SLP-1650+ to remove frequencies above Nyquist.
A combination fixed amplifier and MiniCircuits ZX76-31-SP-S+ attenuator provided tunable gain for this stage, so that the input signal to the ROACH was optimized for its dynamic range~\cite{Patel2014}.
The signal was then passed through a directional coupler, with the coupled output monitored by MiniCircuits ZX47-50LN-S+ power meters.
The mainline coupler output was the stage output, which was connected by a short cable to the ROACH-2 input.

A Raspberry Pi (RPi) microcontroller in the ROACH conditioning stage allowed control and monitoring of the conditioning stage functionality.
Two single-pole double-throw switches controlled the input source, which was manipulated by the RPi GPIO pins:
Both the primary and secondary IF bands from the high frequency mixing stage were connected to the stage.
Additionally, a noise source for calibration of the ROACH was permanently installed and a test tone input was available on the front panel.
The attenuator was controlled by SPI interface, providing \SI{30}{\decibel} of tunable range in discrete sub-dB steps.
One power meter was connected to an Adafruit ADS1115 ADC board, which was regularly read out by the RPi by I$^{2}$C interface.
Prior to any datataking, the power level was verified and the attenuator level adjusted as necessary.

The performance of the ROACH conditioning stage was verified with measurements prior to installation.
The transfer function was measured to be smooth across the signal bandwidth, falling by \SI{5}{\decibel} over the entire range.
The image noise was controlled by the low-pass filter, with the noise power increase limited at the high frequency end to $<$\SI{0.5}{\decibel} by \SI{1.4}{\giga\hertz}.
These characterization measurements also calibrated the power meter response and verified isolation between all input channels.

\begin{figure}
  \centering
  \includegraphics[width=120mm]{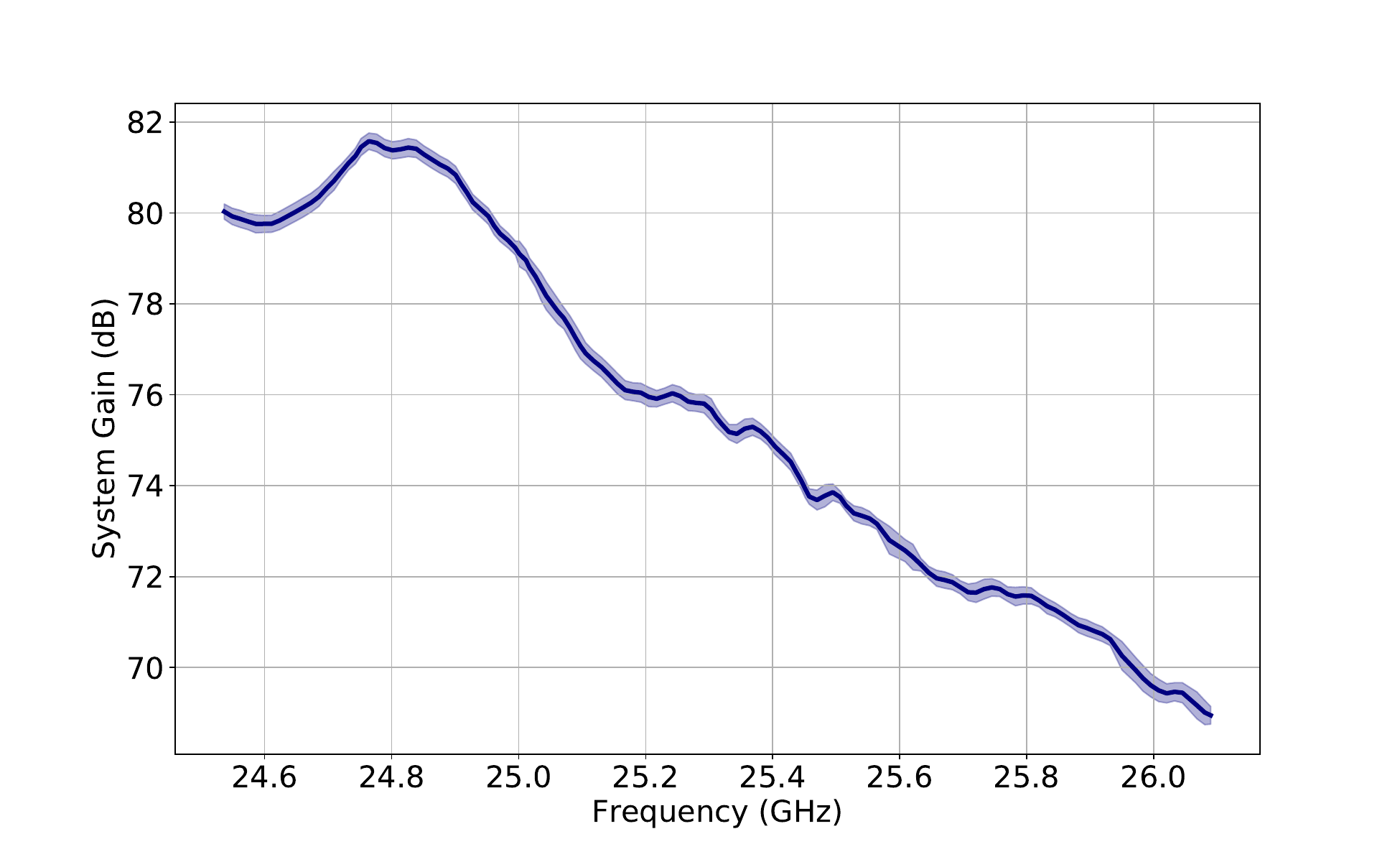}
  \caption{System gain as a function of frequency as determined using the Y-factor method.}
  \label{fig:rf_gain}
\end{figure}

To guarantee the optimal performance of this complex RF system, a Y-factor method was employed to assess the receiver system's RF background and gain characteristics.
This involved changing the terminator temperature and analyzing its effect on the power spectrum using a thermodynamic model.
The results of this analysis are shown in Fig.~\ref{fig:rf_gain}, which illustrates the system gain varying between \SI{69}{dB} and \SI{82}{dB} across the frequency interval from \SI{24.2}{\giga\hertz} to \SI{26.5}{\giga\hertz}.
The noise temperature was determined to be $\sim$\SI{132}{K}, driven primarily by thermal noise from the terminator.
A detailed discussion of this analysis is presented in Appx.~\ref{appx:rf-bkg}.
The frequency-dependence of the RF properties of the system was considered in the final analysis through an extensive electron-data-driven study of how the electron detection efficiency varied with frequency~\cite{PRC}.

\clearpage
\newpage

\section{Gas System}\label{ch5:Gas-system}

The gas system was designed to operate in two configurations: 1) supplying the CRES cell with a controllable pressure of $\mathrm{T_2}$ gas for performing the study of molecular tritium beta decay; and 2) supplying $\mathrm{^{83m}}$Kr, which emits monoenergetic conversion electrons at several energies between \SI{7}{\kilo\electronvolt} and \SI{32}{\kilo\electronvolt}.
The known-energy electrons from $\mathrm{^{83m}}$Kr were used to measure the magnetic field via Eq.~\ref{cres-eq}~\cite{Otten_2008, PRC}, and the K-line electrons at \SI{17.824}{\kilo\electronvolt} \cite{Venos_2018}, close to the tritium endpoint energy of \SI{18.574}{\kilo\electronvolt}, were used to characterize detector response and detection efficiency \cite{PRC} in this region.

\begin{figure}
  \centering
  \includegraphics[origin=c,width=150mm]{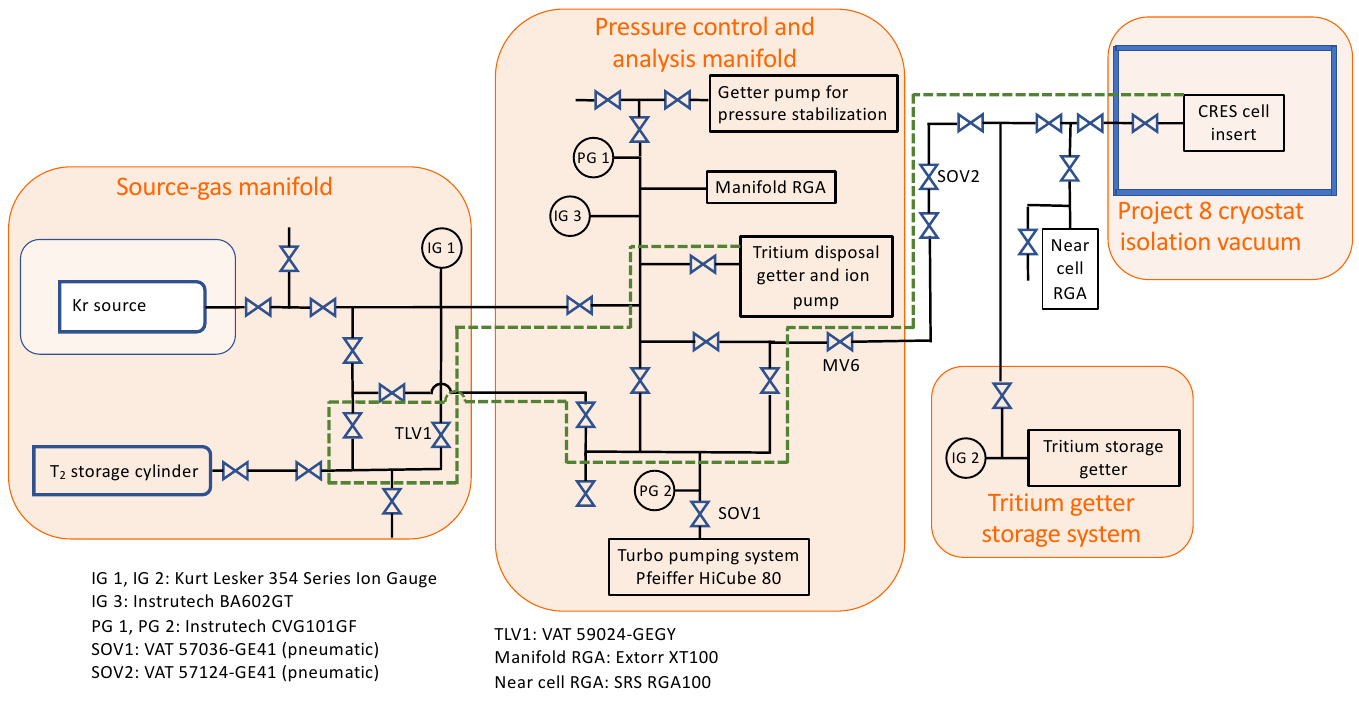}
  \caption{Gas system diagram for the combined $\mathrm{Kr}/\mathrm{T_2}$ gas system.
    The dashed green line shows the path used for pumping $\mathrm{^3He}$ during $\mathrm{T_2}$ data-taking.}
  \label{ch3:gas-system}
\end{figure}

Control of the gas environment was a major consideration in the design of this system. It was necessary to tune the T$_2$ density to maximize the event rate by balancing the rate of decays in the apparatus (driving optimal density higher) with efficient detection (driving optimal density lower).\footnote{While a higher gas density would have increased the number of decays, it also would have resulted in more frequent collisions between the electron and the gas molecules.
This would have reduced the mean time during which electrons could cyclotron-radiate before the first collision, making the signal more difficult to detect \cite{T_ER, Viterbi, deep_learning}.
Even if later parts of the signal could be detected, missing the first portion of the signal would have caused a systematic shift in the measured start energy toward lower values, requiring a correction in the analysis phase \cite{PRC}.} To enable accurate calibration, the $\mathrm{^{83m}}$Kr data needed to be taken in an environment tuned to be as similar in electron-gas collision rate as possible to the conditions during tritium running. In both running configurations, getter pumps were used to irreversibly remove contaminants while reversibly storing and controlling the pressure of hydrogen isotopes: T$_2$ for tritium operation, and H$_2$ for $\mathrm{^{83m}}$Kr operation. The pressure of the relevant hydrogen isotope was adjusted by controlling the appropriate getter's temperature.
Fig.~\ref{ch3:gas-system} shows the valve schematics for the combined $\mathrm{Kr}/\mathrm{T_2}$ gas system, which has three distinct sections: the source-gas manifold, the pressure control and analysis manifold, and the tritium getter storage system.

\begin{figure}
  \centering
  \includegraphics[width=100mm]{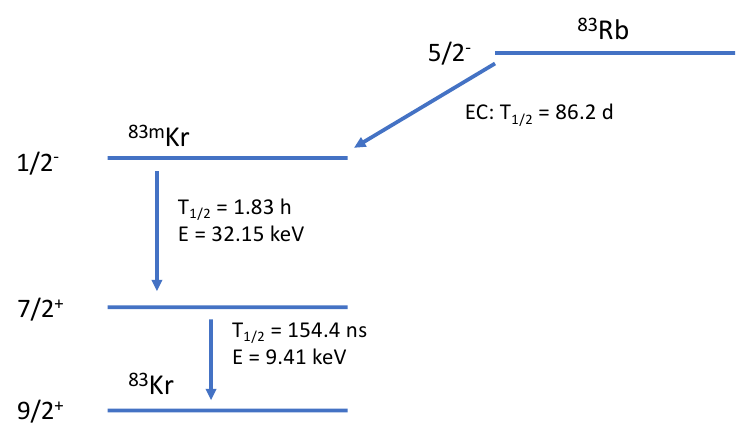}
  \caption{Decay scheme of $\mathrm{^{83}}$Rb.
    $\mathrm{^{83}}$Rb decays to $\mathrm{^{83m}}$Kr through an electron capture process.
    $\mathrm{^{83m}}$Kr decays to its ground state in a cascade of two internal conversion processes, emitting electrons.}
  \label{rubidium-decay-scheme}
\end{figure}

The sources of radioactive gases were attached to the source-gas manifold.
The $\mathrm{^{83m}}$Kr source consisted of $\mathrm{^{83}RbCl}$ adsorbed on Zeolite beads (\SI{2}{\milli\meter} in diameter) encased in lead shielding.
$\mathrm{^{83m}}$Kr was formed as the decay product of $\mathrm{^{83}}$Rb (Fig. \ref{rubidium-decay-scheme}).
As a noble gas, $\mathrm{^{83m}}$Kr was not retained in the zeolite matrix and it diffused into the main gas system \cite{VENOS2005323}.
Two $\mathrm{^{83}}$Rb sources with activities of \SI{8}{\milli\curie} and \SI{6}{\milli\curie}, installed in July 2018 and July 2019 respectively, were used for the majority of the calibration studies.

\begin{figure}
  \centering
  \includegraphics[width=100mm]{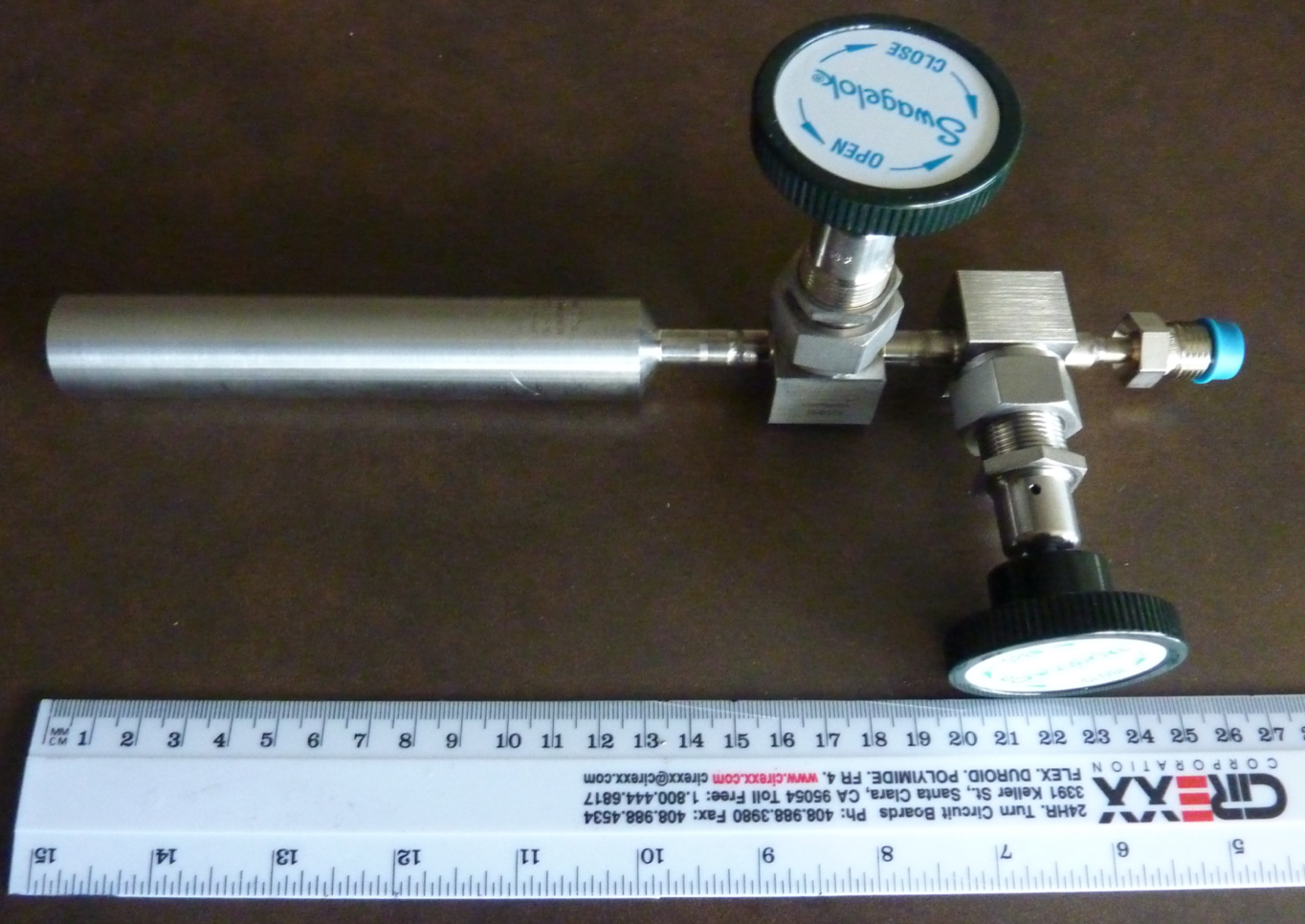}
  \caption{The tritium storage cylinder.
  }
\label{ch3:t2-cylinder}
\end{figure}

A stainless steel Swagelok SS-4CS-TW-50 cylinder was used to store \SI{2}{\curie} of $\mathrm{T_2}$ gas from American Radiolabeled Chemicals (Fig.~\ref{ch3:t2-cylinder}).
Two Swagelok SS-4-BG-TW VCR all-metal valves were welded to the cylinder with a small volume enclosed between them to allow for a controlled release of the gas into the system (by successive opening and closing).
To further control the release of tritium gas into the system, a digitally-controlled VAT 59024-GEGY leak valve (TLV1 in Fig.~\ref{ch3:gas-system}) was included in the source-gas manifold.

The pressure control and analysis manifold contained the measurement devices and pumps.
Two Instrutech CVG101 convection gauges measured pressures in the range of \SIrange[range-units = single]{1e-4}{1}{\torr}.
An Instrutech BA602 ion gauge measured low pressures down to \SI{4e-10}{\torr}.
The gas composition was assessed using an ExTorr XT100 residual gas analyzer (referred to as the `manifold RGA' from this point forward).

The main turbopump, a Pfeiffer HiCube 80, was connected to the system via a VAT 57036-GE41 solenoid valve (SOV1 in Fig.~\ref{ch3:gas-system}), which automatically shut during power fluctuations to prevent radioactive gas leakage into the laboratory.
Base pressures of $\sim$ \SI{e-9}{\torr} were obtained using the HiCube pump after the system's bake-out.
To monitor for tritium leaks, radioactive contamination surveys using swipes assayed with a liquid scintillation counter were done weekly from multiple locations on the gas system. The system was also monitored for leaks by performing regular automated scans with the RGA to detect any incursion of atmospheric gases (e.g., argon).

In addition to the mechanical turbomolecular pump, two non-evaporable getters were included in this manifold.
An SAES NEXTorr D 100-5 NEG pumping system contained the disposal getter, which was used to reduce the tritium partial pressure in the system to a negligible level before $\mathrm{^{83m}}$Kr calibration runs and after the completion of data-taking with tritium (Tbl.~\ref{tbl:getters}).
The integrated diode pump made it possible to also remove noble gases that could not be pumped by the getter.

A SAES GP-50 non-evaporable getter pump was used to control the density of hydrogen gas during $\mathrm{^{83m}}$Kr measurements to match the scattering rate under tritium running conditions (Tbl.~\ref{tbl:getters}).
The pressure in the gas system was an easily measurable proxy that correlated well to the electron scattering rate in the CRES cell, so it was used as the process variable.\footnote{It was an imperfect but still usable proxy, in the following ways. 1) The gas pressure and density in cold cell and at the warm ion gauge were not the same, but were monotonically related in a time-stable way. Also, 2) the ion-gauge-measured pressure and electron scattering rate depend differently on gas composition, making it necessary to measure their relationship separately for the T$_2$ data and the $\mathrm{^{83m}}$Kr data; however, this was done successfully and pressures were set that closely matched the electron scattering rates in the two run modes~\cite{PRC}.}
The target pressure was settable over a range of several orders of magnitude and maintained by an OMEGA CN16DPt PID controller unit which adjusted the electrical current supplied to the activation filament in the non-evaporable getter pump.

Similarly, the tritium getter storage system was designed to regulate the tritium gas pressure in the CRES cell.
This system contained a SAES St-172 zirconium-based non-evaporable getter (referred to as the tritium control getter from this point forward) that was loaded with molecular tritium. 
A dedicated Omega CN16DPt PID controller regulated the pressure of tritium inside the CRES cell by temperature-controlling the tritium control getter.
This getter also pumped background gases in the CRES cell (Tbl.~\ref{tbl:getters}).

\begin{table}
\begin{center}
\begin{tabular}{ | c | c | m{6cm} | } 
  \hline
  Getter name & \shortstack{Manufacturer \& model number} & Function \\ 
  \hline
  \shortstack{Hydrogen control \\ getter} & SAES GP-50 & Raising the gas pressure for the $\mathrm{^{83m}}$Kr runs by introducing hydrogen into the system in order to reproduce the tritium run conditions. \\ 
  \hline
  \shortstack{Disposal getter} & SAES NEXTorr D 100-5 NEG & Maintaining low background pressures for the $\mathrm{^{83m}}$Kr runs after the system was tritiated and absorbing the tritium after the conclusion of the tritium runs. \\ 
  \hline
  \shortstack{Tritium control \\ getter} & SAES St-172 & Regulating the tritium pressure and pumping the background gases in the CRES cell during tritium runs. \\ 
  \hline
\end{tabular}
\caption{Getter pumps used in the gas system and their functions.}
\label{tbl:getters}
\end{center}
\end{table}

\begin{figure}
  \centering
  \includegraphics[width=120mm]{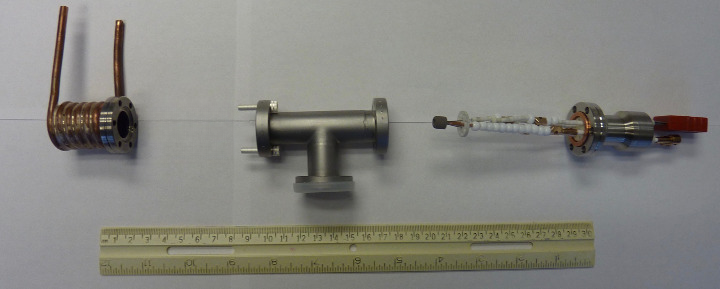}
  \caption{Tritium getter storage system components before before their final assembly.
  }
  \label{ch3:getter-compartment}
\end{figure}

The tritium control getter was installed in a mini-conflat half nipple.
Around the half nipple a copper tube was brazed, through which cooling water was flowed while the getter was running at high temperatures.
The delicate filament connectors of the getter were crimped into two small copper tubes which were connected to a vacuum feed-through.
A C-type thermocouple was mounted inside the annular storage getter for measuring its temperature.
The thermocouple connectors were electrically isolated with ceramic beads.
Fig.~\ref{ch3:getter-compartment} shows this setup before the final assembly.
A Kurt J. Lesker (KJL) 354 series ion gauge was used to monitor the pressure in the tritium getter storage system, providing the reference signal for the PID pressure controller.
When not taking tritium data, the tritium was confined to the tritium storage getter region by closing a manual valve. The tritium getter storage system was connected to the rest of the gas system via a shut-off valve (SOV2 in Fig.~\ref{ch3:gas-system}); this valve would automatically close to limit the tritium-exposed volume in case of a power outage.
A Stanford Research Systems RGA100 residual gas analyzer (referred to as the `near-cell RGA' from this point forward) was installed as close as possible to the CRES cell to assess gas composition during tritium data-taking (Fig.~\ref{ch3:gas-system}).

The gas-pressure-control system worked as intended in both $\mathrm{^{83m}}$Kr calibration and tritium running modes.
Pressures in the range \SIrange[range-units = single]{1.2}{2.0}{\micro\torr} were used for the main data sets\footnote{The ion gauges were set to H2-equivalent, so the pressures are different for $\mathrm{^{83m}}$Kr and tritium data to account for the ion gauge's different sensitivity to hydrogen isotopes and Kr.}, with $\pm3\%$ stability run-to-run. With this system in place, CRES data were acquired in $\mathrm{^{83m}}$Kr calibration mode at a range of pressures to characterize the dependence of detection efficiency on gas pressure. As described in Ref.~\cite{PRC}, this information was used to inform the choice of set pressure for the tritium data run to maximize event rate.

\subsection{T$_2$ Data-taking}
To test the storage-getter-based density control mechanism without the radioactivity-associated safety concerns of working with tritium, tests using deuterium were performed first.
The relation between the equilibrium pressure of hydrogen isotopes and the getter temperature was examined.
The results of these tests are described in detail in Appx.~\ref{appx:d2-test}.

After the conclusion of the deuterium tests, the tritium cylinder was installed.
The tritium was absorbed into the tritium control getter, after which $\mathrm{^3He}$ (the product of tritium decays that had occurred before installation) was removed with the ion pump.

\begin{figure}
  \centering
  \includegraphics[width=120mm]{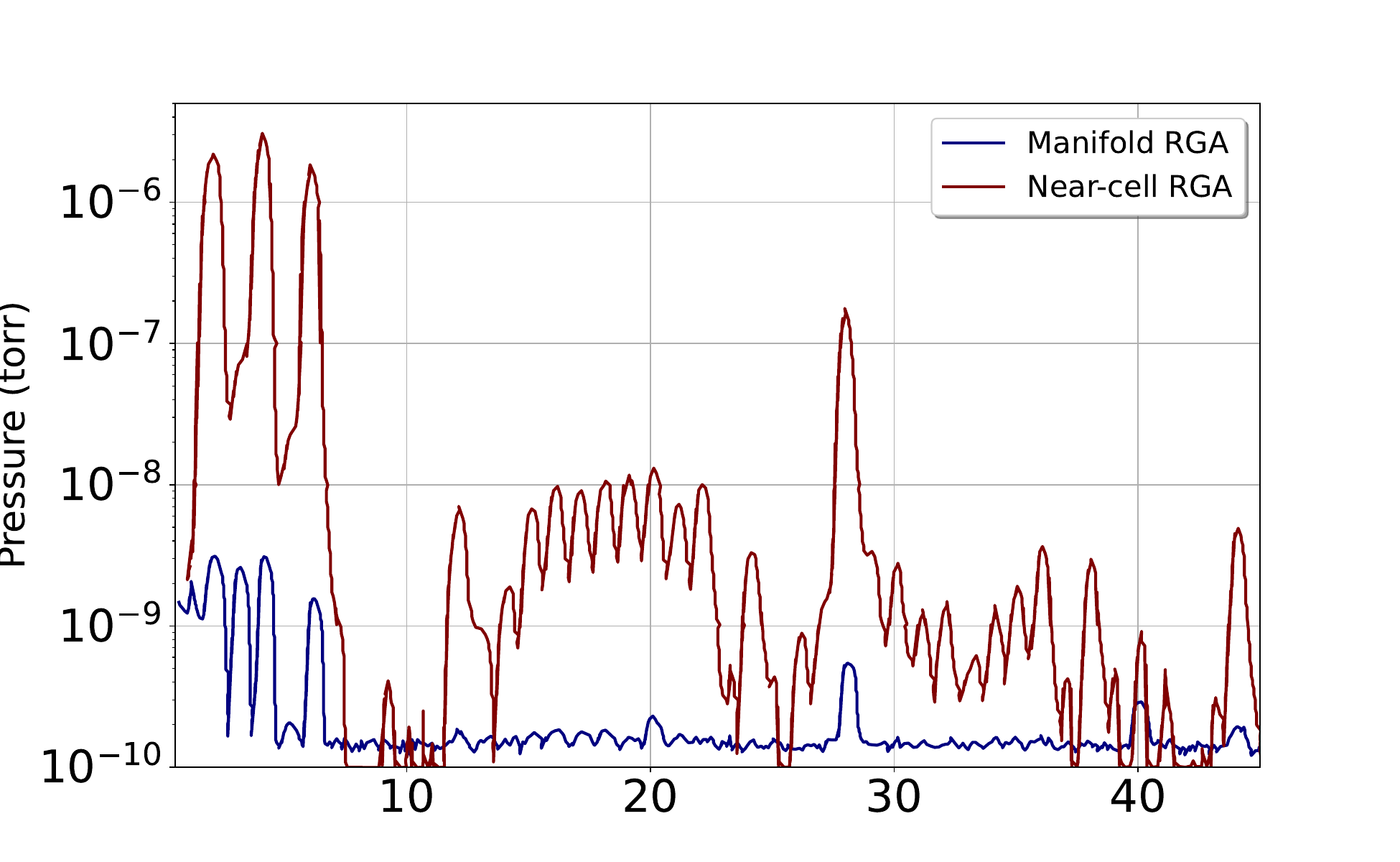}
  \caption{The gas composition measured by the manifold RGA (top) and the near-cell RGA (bottom) during tritium data taking.
    The two spectra are approximate measures of the partial pressures in the ion pump and in the CRES cell.
  }
  \label{ch3:RGA}
\end{figure}

One challenge in maintaining high tritium purity in the CRES cell was the presence of $\mathrm{^3He}$, which came from two sources.
1) The tritium adsorbed on the stainless-steel walls of the getter compartment during the tritium fill released $\mathrm{^3He}$ into the system.
This source was suppressed by baking the getter compartment for several hours, which drove the tritium into the tritium control getter.
2) $\mathrm{^3He}$ was also produced in tritium decay inside the tritium control getter, where it was not all trapped.
After the impact of this effect was discovered, the gas system was modified to establish a path for continuous pumping by the ion pump (Fig. \ref{ch3:gas-system}).
The gas flow through the leak valve was tuned to efficiently remove $\mathrm{^3He}$ from the gas system without excessive loss of the $\mathrm{T_2}$ inventory in the tritium control getter.
The manifold RGA, which was placed close to the ion pump, was used to estimate an upper limit of a tritium pumping rate of \SI{34}{\milli\curie / \year}.
Fig.~\ref{ch3:RGA} compares the gas composition measured by the two residual gas analyzers during the tritium data campaign confirming that the continuous pumping worked as expected.
The near-cell RGA's spectrum was dominated by mass 2, 4, and 6, corresponding to $\mathrm{H_2}$, $\mathrm{HT}$, and $\mathrm{T_2}$.
This demonstrated the presence of high partial pressures of hydrogen isotopes in the CRES cell.
Smaller amounts of $\mathrm{^3He}$ and $\mathrm{CO}$ (mass 3 and 28) were also apparent as the major background gases.
The $\mathrm{^3He}$ was more prominent in the manifold RGA's spectrum both because the conductance of the leak valve (TLV1) was higher for this light isotope, and because it was only removed by the ion pumping.
A quantitative analysis of the gas composition in running conditions, based on residual gas analyzer data, is described in~\cite{PRC}. In brief, it was found that 91$\pm$5\% of scattering events during tritium data-taking were from hydrogen isotopes, demonstrating the gas system's good performance.

\section{Conclusion}
Cyclotron Radiation Emission Spectroscopy (CRES) has been demonstrated as a novel technique for measuring the energy of electrons emitted in beta decay, including the endpoint spectrum of tritium decay.
In this work we detailed the apparatus used in Phase II of the Project 8 experiment, which was used to set the first frequency-based neutrino mass limit \cite{PRL,PRC}.
Improvements in Phase II over the previous CRES apparatus included the use of a more efficient circular waveguide geometry, an improved RF detection chain, a more detailed characterization of the RF background, and a system for delivering pressure-controlled radioactive gases, including the tritium that enables CRES to be used for a neutrino mass measurement.

\section{Acknowledgment}
This material is based upon work supported by the following sources: the U.S. Department of Energy Office of Science, Office of Nuclear Physics, under Award No.~DE-SC0020433 to Case Western Reserve University (CWRU), under Award No.~DE-SC0011091 to the Massachusetts Institute of Technology (MIT), under the Early Career Research Program to Pacific Northwest National Laboratory (PNNL), a multiprogram national laboratory operated by Battelle for the U.S. Department of Energy under Contract No.~DE-AC05-76RL01830, under Early Career Award No.~DE-SC0019088 to Pennsylvania State University, under Award No.~DE-FG02-97ER41020 to the University of Washington, and under Award No.~DE-SC0012654 to Yale University; the National Science Foundation under Grant No.~PHY-2209530 to Indiana University, and under Grant No.~PHY-2110569 to MIT; the Cluster of Excellence “Precision Physics, Fundamental Interactions, and Structure of Matter” (PRISMA+ EXC 2118/1) funded by the German Research Foundation (DFG) within the German Excellence Strategy (Project ID 39083149); the Karlsruhe Institute of Technology (KIT) Center Elementary Particle and Astroparticle Physics (KCETA); the MIT Wade Fellowship; the LDRD Program at PNNL; the University of Washington Royalty Research Foundation; and Yale University.  A portion of the research was performed using Research Computing at PNNL.  The isotope(s) used in this research were supplied by the United States Department of Energy Office of Science by the Isotope Program in the Office of Nuclear Physics.

\begin{appendices}
\section{Investigation of the RF Noise Background}\label{appx:rf-bkg}

In this section, we present an analysis of the noise and the gain of the complex receiver chain of Phase II.
To this end, the terminator temperature was altered and its effect on the power spectrum was analyzed using a thermodynamic model.

The RF background was determined from the power spectral density (PSD) when no electrons were trapped in the cell. 
The background was measured using the ROACH-2 digitizer with a frequency span of \SI{24.5}{GHz} to \SI{26.1}{GHz}. 
The spectrum was obtained from applying the Welch method to 26,214,400 time samples~\cite{welch}.

The RF background was measured as a function of cell temperature from \SI{58}{K} to \SI{110}{K}.
The cell and amplifier temperatures were monitored using calibrated Lakeshore Cernox-87821 temperature sensors.
Other temperatures along the cryogenic insert were measured with PT100 sensors.

A thermodynamic model was developed to describe the RF background as the sum of the black-body radiation of each mechanical component of the cryogenic insert and the Johnson noise of each electronic component.
Recall that in the high-temperature regime, the black-body radiation is $P_n/B = k_{\mathrm{B}} T$, where $P_n$ is the noise power, $B$ is the total bandwidth over which the noise is measured, $k_{\mathrm{B}}$ is the Boltzmann constant, and $T$ is the temperature.
For the noise contribution of electronic components, the Johnson noise is expressed in the same form using an equivalent noise temperature $T_n$.
From the measured noise power, an equivalent system noise temperature, $T_\mathrm{sys}$, is defined,
\begin{equation}
  P_n = T_\mathrm{sys}G k_{\mathrm{B}} B,
  \label{eqn:noise_power}
\end{equation} 
where G is the total system gain.

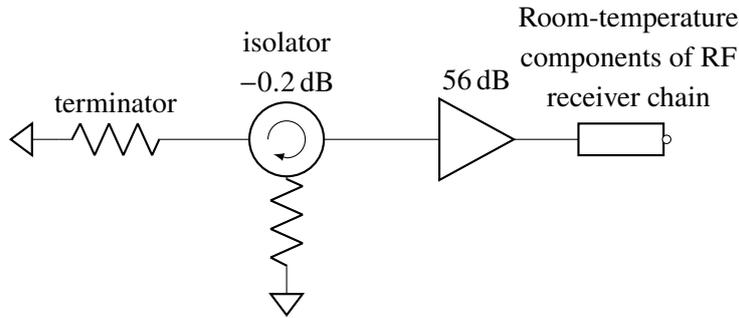
\begin{figure}[ht]
  \centering
  \begin{circuitikz}
    \draw(0,0)node[sground, rotate = -90]{} 
    to[R=terminator] (1.5,0)--(2.5,0)
    (3,0)node[circulator, align = left]{}(3.5,0)--(5,0)
    to [amp, l=\SI{56}{dB}] (6,0)--(6.8,0)
    to [generic, -o](8,0);
    \draw(3,-1.7)node[sground]{} 
    to[R](3,-0.5);
    
    \draw(3,0.5)node[anchor=south, align=center]{isolator\\ \SI{-0.2}{dB}};
    \draw(7.5,0.3)node[anchor=south, align=center]{Room-temperature\\ components of RF\\ receiver chain };
 
  \end{circuitikz}
  \caption{Thermal model for the noise power of the cryogenic insert. 
  The black-body radiation from the terminator traveled up the waveguide, through the isolator, and into the first-stage amplifier.
  The isolator absorbed some of the terminator's radiation and re-emitted it as its own black-body radiation.}
  \label{fig:thermodynamic_model}
\end{figure}

The thermodynamic model in Fig.~\ref{fig:thermodynamic_model} is developed to describe the noise power in Phase II. 
The fraction of thermal photons from the terminator that reaches the amplifier is $\eta_\mathrm{iso}$, the isolator's transmissivity. 
The isolator absorbs a fraction ($1-\eta_\mathrm{iso}$) of photons and emits blackbody radiation $P/B = \epsilon_\mathrm{iso}k_{\mathrm{B}} T_\mathrm{iso}$, where $T_\mathrm{iso}$ and $\epsilon_\mathrm{iso}$ are the isolator's temperature and emissivity. 
By Kirchoff's law, the isolator's emissivity is equal to the fraction of the blackbody radiation that the isolator absorbs, $\epsilon_\mathrm{iso} = 1-\eta_\mathrm{iso}$. 
The amplifier adds to the signal its Johnson noise, which is written as an equivalent noise temperature $T_\mathrm{amp}$. 

The noise contribution from the next components of the receiver system is negligible except for the contribution from digitization noise and image noise. 
Hence, $T_\mathrm{amp}$ is considered to be the equivalent noise temperature of the receiver, from the first-stage amplifier to the digitizer.

\begin{figure}
  \centering
  \includegraphics[width=120mm]{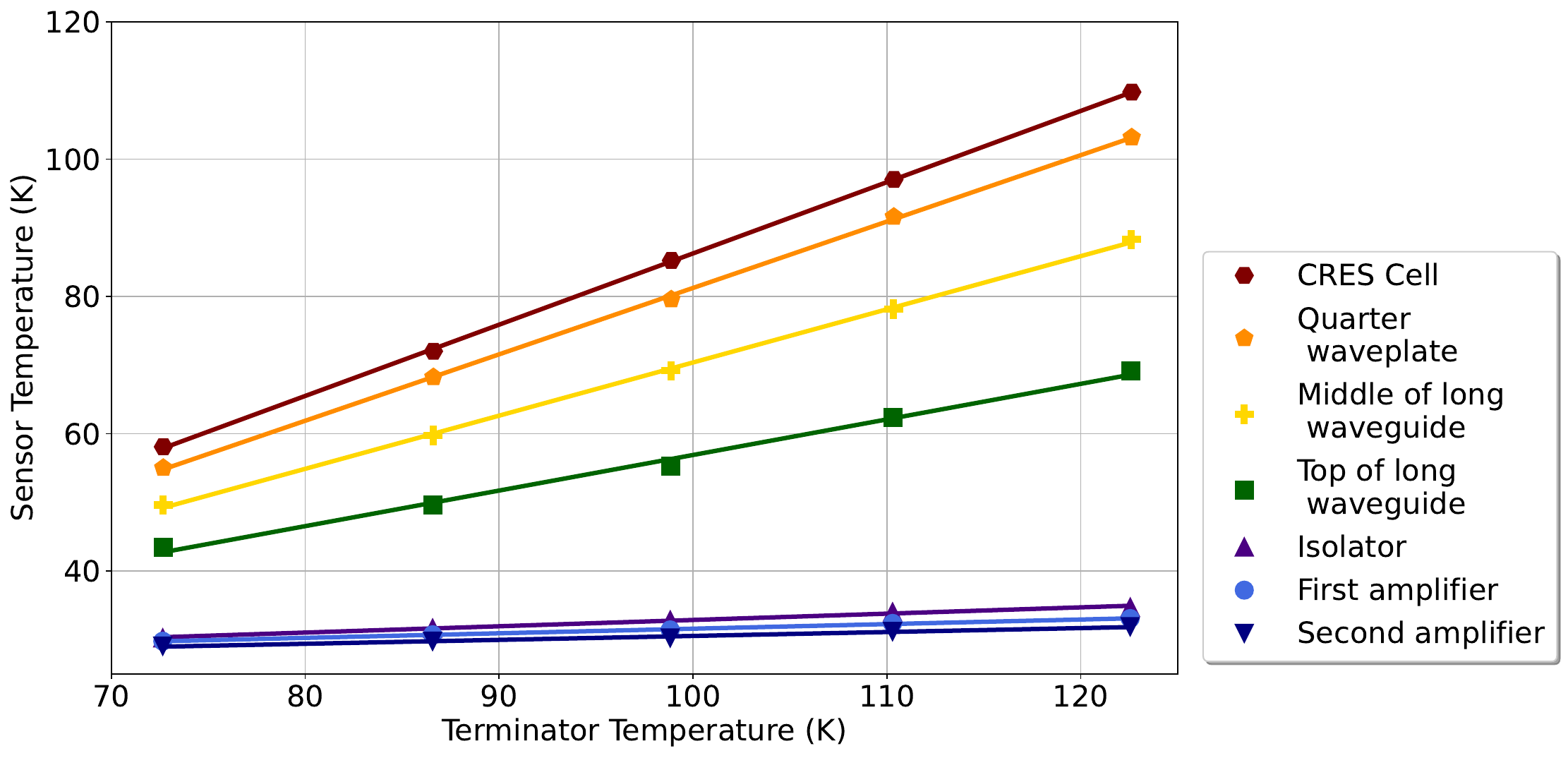}
  \caption{Temperature along the cryogenic insert as a function of terminator temperature with the linear fits.
    The plot proves the linear dependence of the temperatures along the cryogenic insert on the terminator temperature.
    The plot is also a testimony to the effectiveness of the thermal break in maintaining low temperatures for the isolator and the amplifiers.}
  \label{fig:temp_sensors}
\end{figure}

The resulting noise temperature from the thermodynamic model is
\begin{eqnarray}
  \begin{aligned}
T_\mathrm{sys} = T_\mathrm{term} \eta_\mathrm{iso} + T_\mathrm{iso}(1- \eta_\mathrm{iso}) +  T_\mathrm{amp}.
  \end{aligned}
  \label{eqn:tsys}
\end{eqnarray}
The isolator temperature $T_\mathrm{iso}$ depends linearly on $T_\mathrm{term}$ and is written as $T_\mathrm{iso} = aT_\mathrm{term} + b$. 
Sensor data were used to extract the parameters a and b (Fig.~\ref{fig:temp_sensors}). 
$T_\mathrm{sys}$ is written as a linear function of $T_\mathrm{term}$, $T_\mathrm{sys}$ and the RF background is modeled as

\begin{eqnarray}
	\begin{aligned}
		P_n = G B k_{\mathrm{B}} ( [ \eta_\mathrm{iso} + a(1-\eta_\mathrm{iso})] T_\mathrm{term} \\ + b(1-\eta_\mathrm{iso}) + T_\mathrm{amp} )
	\end{aligned}
	\label{eqn:thermal_model}
\end{eqnarray}
and the linear dependence on $T_\mathrm{term}$ is made explicit.
Isolator transmissivity $\eta_\mathrm{iso} = 0.955 \pm 0.045$ is used for the isolator transparency and its considerable uncertainty. 

The rectangular waveguide and CRES cell are treated as lossless components in this model. 
However, the effect of the quarter-wave plate is non-negligible.
With the inclusion of the quarter-wave plate in the thermal model, Eq.~\ref{eqn:tsys} is modified to
\begin{eqnarray}
  \begin{aligned}
		T_\mathrm{sys} = T_{\mathrm{term}} \eta_\mathrm{qwp} \eta_\mathrm{iso} + T_\mathrm{qwp}(1- \eta_\mathrm{qwp})\eta_\mathrm{iso} + \\T_\mathrm{iso}(1- \eta_{\mathrm{iso}}) +  T_\mathrm{amp},
  \end{aligned}
  \label{eqn:tsys_qwp}
\end{eqnarray}
where $T_\mathrm{qwp}$ ($\eta_\mathrm{qwp}$) is the temperature (transparency) of the quarter-wave plate.
Assuming that the quarter-wave plate temperature was close enough to the terminator temperature, Eq.~\ref{eqn:tsys_qwp} should only be more accurate than Eq.~\ref{eqn:tsys} by a few percent.
The model uncertainty is estimated to be 4\% by taking the thermal photon's temperature as a weighted average of the terminator and quarter-wave plate and assuming \SI{1}{\decibel} insertion loss for the quarter-wave plate.

The amplifier noise and gain were fitted from measurements of the RF background at various terminator temperatures.
From the linear relationship in Eq.~\ref{eqn:thermal_model}, the terminator temperature, $x_\mathrm{int}$, that resulted in zero noise power, $P_n = 0$, was determined regardless of the complicated frequency-dependent gain.
The amplifier noise temperature was deduced, where
\begin{eqnarray}
  \begin{aligned}
		T_\mathrm{amp} = -[ \eta_\mathrm{iso} + a(1-\eta_\mathrm{iso})] x_\mathrm{int} \\- b(1-\eta_\mathrm{iso}).
  \end{aligned}
  \label{eqn:ta}
\end{eqnarray}
The gain was evaluated from the slope of the line-fit $m$, 
\begin{eqnarray}
  \begin{aligned}
		G = \frac{m}{B k_{\mathrm{B}}}\frac{1}{\eta_\mathrm{iso} + a \left(1-\eta_\mathrm{iso}\right)}.
  \end{aligned}
  \label{eqn:G}
\end{eqnarray}
The calculated amplifier noise temperature was used to find the system noise temperature for a given terminator temperature using Eq.~\ref{eqn:tsys}.

\begin{figure}
  \centering
  \includegraphics[width=120mm]{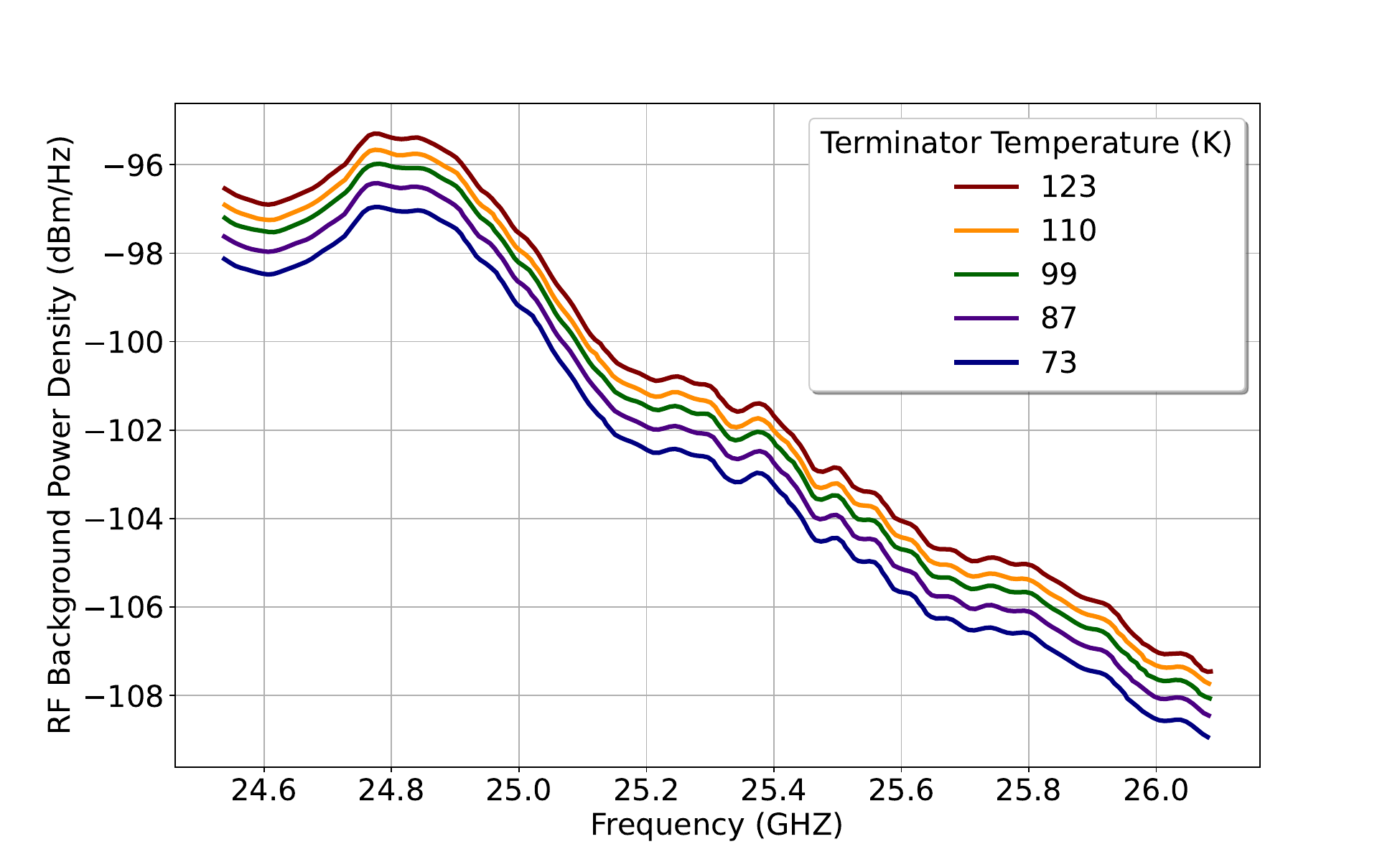}
  \caption{RF background power spectral density as a function of frequency and terminator temperature.}
  \label{fig:rf_background}
\end{figure}

\begin{figure}
  \centering
  \includegraphics[width=120mm]{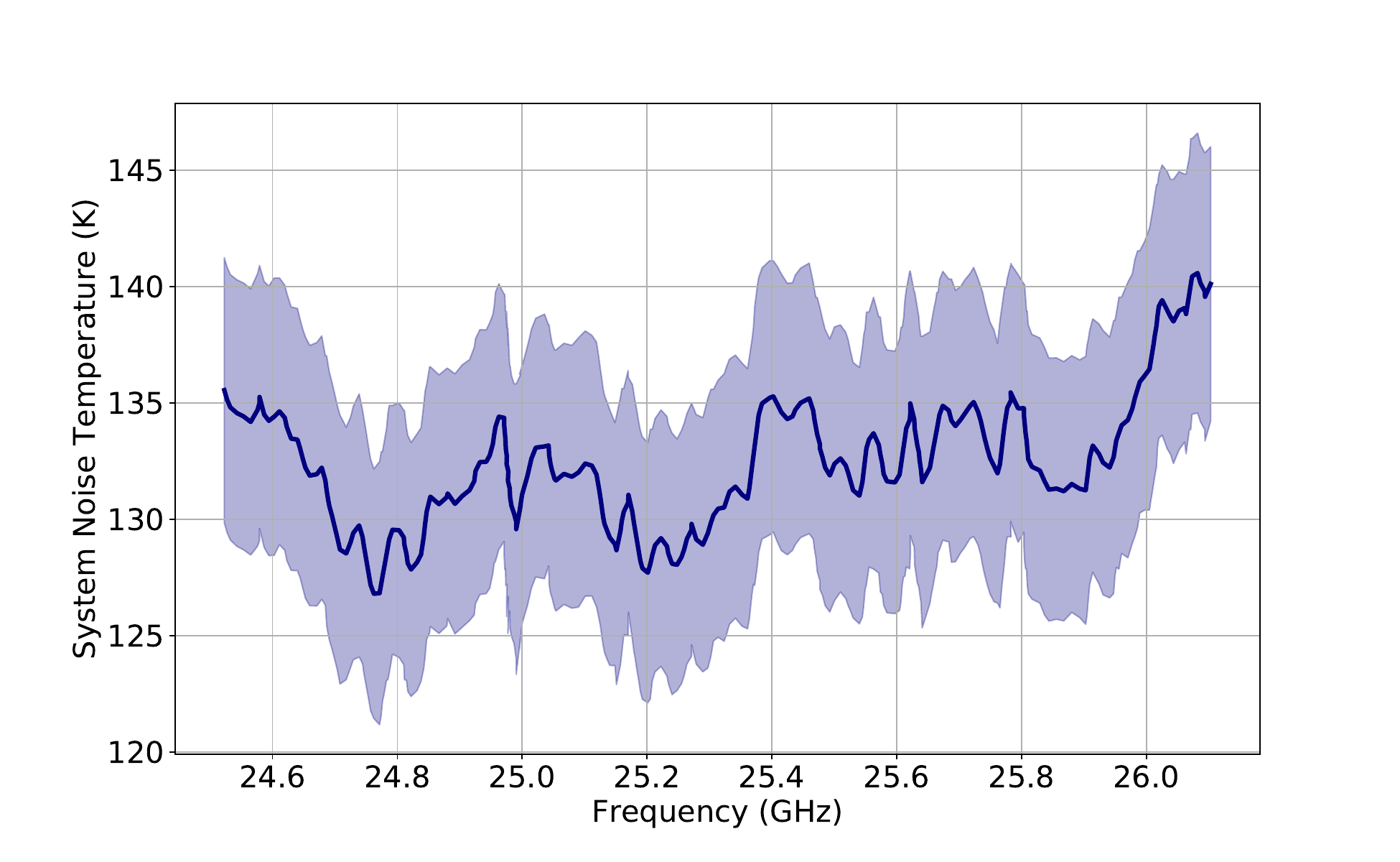}
  \caption{System noise temperature obtained from the Y-factor method when the terminator was at \SI{99}{\kelvin} and the cell was at \SI{85}{\kelvin}.}
  \label{fig:tsys}
\end{figure}

Fig.~\ref{fig:rf_background} shows the RF background as a function of terminator temperature.
The system noise temperature was calculated when operating the CRES cell at \SI{85}{K} (Fig.~\ref{fig:tsys}).
The system noise temperature was estimated to be \SI{132}{K} with a 5\% uncertainty. 

Finally, the system gain was determined by using Eq.~\ref{eqn:G} (see Figure~\ref{fig:rf_gain}).
The gain ranged from \SI{82}{dB} to \SI{70}{dB} with an average uncertainty (in the logarithmic scale) of about 0.3\%.

\section{$\mathrm{D_2}$ Tests}\label{appx:d2-test}

Tests using deuterium were performed to investigate the performance of the storage getter without the safety concerns associated with tritium operation.
One of the objectives for these tests was to examine the relation between the equilibrium pressure and the getter temperature to ensure the proper functionality of the tritium control getter.
For hydrogen isotopes in a metallic getter, the equilibrium pressure $P$ (in \si{\torr}) can be derived from Sievert's law, in which
\begin{equation}\label{ch3:sievert-eq}
    \log{P} = a + 2 \log{q} - \frac{b}{T},
\end{equation}
where $q$ is the hydrogen concentration in the metallic alloy (in \si{\liter\cdot\torr/\gram}), $T$ is the temperature (in \si{\kelvin}), and $a$ and $b$ are constants that must be measured for each getter individually.
According to this equation, the equilibrium pressure increases with the getter's temperature and hydrogen concentration.

\begin{figure}[ht]
  \centering
  \includegraphics[width=120mm]{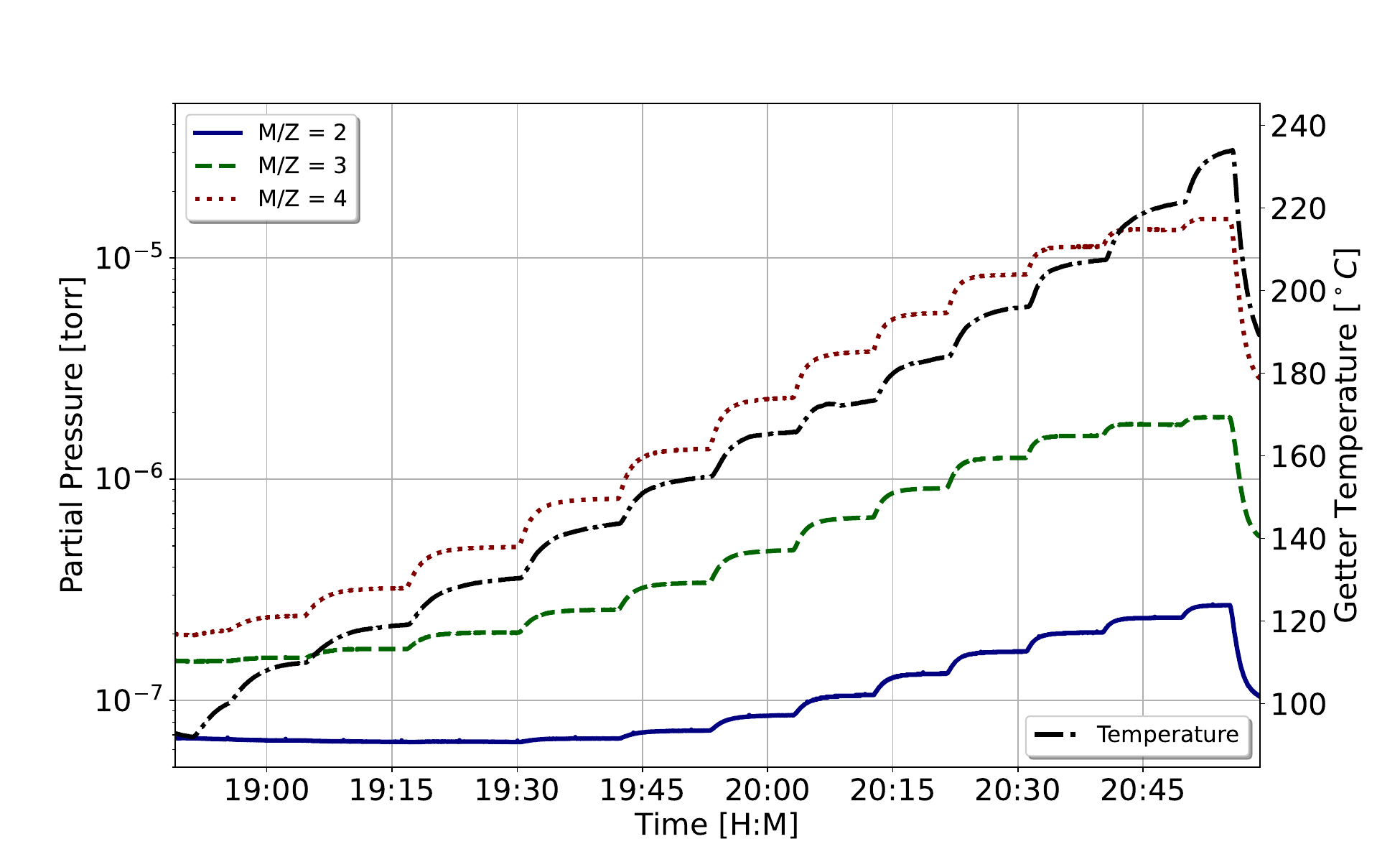}
  \caption{Measured partial pressures of hydrogen isotopes with time as the getter temperature was changed.
  }
  \label{ch3:d2-test}
\end{figure}

The tritium control getter was loaded with deuterium at its operating temperature of \SI{400}{\degreeCelsius} until an equilibrium pressure of \SI{2.2e-6}{\torr} was achieved.
Next, the tritium control getter was cooled down to room temperature and then the filament's current was incremented in \SI{0.5}{\ampere} steps.
The partial pressures of hydrogen isotopes measured with the residual gas analyzer are plotted in Fig.~\ref{ch3:d2-test}.
The main gas component was $\mathrm{D_2}$ with smaller partial pressures of $\mathrm{HD}$ and $\mathrm{H_2}$ present in the system. 
Although a C-type thermocouple was mounted touching the interior surface of the annular getter, its thermal coupling to the surface of the getter was unknown. The temperature was therefore monitored by tracking the resistance of the molybdenum filament inside the getter instead.

The ion gauge measurement was used to find the equilibrium pressure for each step with a given current.
Note that the residual gas analyzer data could only be used for relative pressure measurement due to the unknown normalization scheme.
\begin{figure}
  \centering
  \includegraphics[width=120mm]{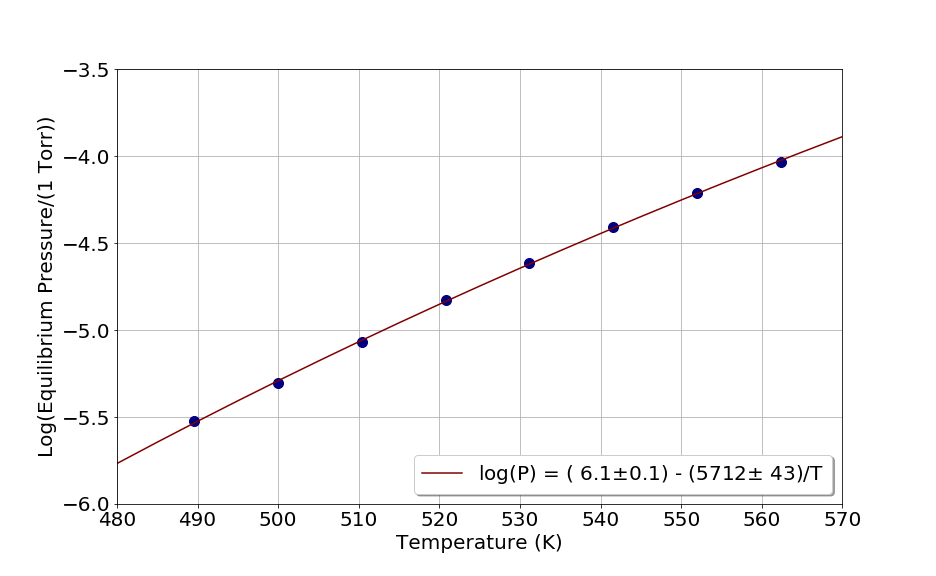}
  \caption{The measured equilibrium pressure vs. temperature for eight different values of the getter's filament current.
    The data were fitted with the Sievert's law equation (Eq. \eqref{ch3:sievert-eq}).}
  \label{ch3:sievert-plot}
\end{figure}

Fig.~\ref{ch3:sievert-plot} shows the equilibrium pressure in the system as a function of the getter temperature.
The data were fitted with Sievert's law, Eq. \eqref{ch3:sievert-eq}, to deduce the empirical coefficients.
According to the SAES St 172 getter manual, the value for the temperature coefficient (b in Eq.\eqref{ch3:sievert-eq}) is \SI{5730}{\kelvin} which is in agreement with our extracted value of \SI{5712+-43}{\kelvin}.
The total amount of deuterium in the getter was also estimated to be \SI{4.4+-0.5}{\torr\liter} using the fit in Fig.~\ref{ch3:sievert-plot}, Eq.\eqref{ch3:sievert-eq}, the getter mass (550 mg), the value of $a=4.45$ for hydrogen from the manual \cite{getter}, and the fact that the equilibrium pressure of a hydrogen isotope is inversely proportional to the square root of the atomic mass number \cite{D2-pressure}.

\end{appendices}

\bibliographystyle{iopart-num}
\bibliography{main}

\end{document}